\documentclass[12pt,amsmath,amssymb,superscriptaddress,floatfix,showpacs]{iopart}

\expandafter\let\csname equation*\endcsname\relax
\expandafter\let\csname endequation*\endcsname\relax

\usepackage{graphicx}
\usepackage{mathdots}
\usepackage{epstopdf} 
\usepackage[usenames,dvipsnames]{color}
\usepackage{bm}

\usepackage{inputenc}
\usepackage{xcolor}
\usepackage{tikz}
\usepackage{amsmath}
\usepackage{amsfonts}

\usepackage{booktabs}

\usepackage[colorlinks,bookmarks=false,citecolor=blue,linkcolor=red,urlcolor=blue]{hyperref}
\usepackage[colorlinks,bookmarks=false,citecolor=blue,linkcolor=red,urlcolor=blue]{hyperref}
\usepackage{xcolor}
\usepackage{mathdots}
\usepackage{float}
\usepackage{needspace}
\usepackage{soul}
\usepackage[usenames,dvipsnames]{color}

\newcommand{\euler}{\mathrm{e}}
\newcommand{\iu}{\mathrm{i}}

\begin{document}
\title{Line defects in infinite networks of resistors}

\def\correspondingauthor{\footnote{Corresponding author: gabor.szechenyi@ttk.elte.hu}}

\author{R\'obert N\'emeth$^{1}$, J\'ozsef Cserti$^1$, G\'abor Sz\'echenyi$^{2,3}$\correspondingauthor}
\address{$^1$ 	Department of Physics of Complex Systems, ELTE E\"otv\"os Lor\'and University, H-1117 Budapest, Hungary  \\ $^2$ Department of Materials Physics, ELTE E\"otv\"os Lor\'and University, H-1117 Budapest, Hungary\\ $^3$ HUN-REN Wigner Research Centre for Physics, H-1525 Budapest, Hungary}




\begin{abstract}
We study infinite resistor networks perturbed by line defects, in which the resistances are periodically modified along a single line. Using the Sherman–Morrison identity applied to the reciprocal-space representation of the lattice Green’s function, we develop a general analytical framework for computing the equivalent resistance between arbitrary nodes. The resulting expression is a one-dimensional integral that is evaluated exactly in special cases. While our analysis is carried out for the square lattice, the method readily extends to other lattice geometries and networks with general impedances. Therefore, this framework is useful for studying the boundary behavior of topolectrical circuits, which serve as classical analogs of topological insulators.

\end{abstract}

\maketitle

\section{Introduction}

Networks of identical resistors arranged in regular lattices have long served as foundational models in both theoretical and applied physics. Among these, the infinite square lattice—where each edge corresponds to an identical resistance $R$—stands out as one of the most extensively studied configurations. The classical problem of determining the equivalent resistance between two arbitrary nodes in such a lattice was first solved in the 1950s \cite{vanderpol1950operational}, with its simplest case—the resistance between two nearest neighbors, equal to $R/2$—now widely recognized \cite{10.1119/1.1970777, Gnädig_Honyek_Riley_2001}. This result paved the way for a broader analytical framework using lattice Green’s functions \cite{10.1119/1.1285881, PhysRevE.108.044148}, making it possible to calculate the resistance between any two nodes in an infinite, periodically tiled space \cite{Cserti_2011}. This formalism has since been applied to various two- and three-dimensional lattice geometries \cite{OWAIDAT20191621, Asad2013, Tan2022}, supporting both exact and asymptotic evaluations of effective resistances \cite{Essam_2009, PhysRevE.82.011125}.

Beyond their intrinsic mathematical appeal, resistor networks are closely connected to various areas of physics and applied mathematics. They serve as discrete analogs of random walk problems, where effective resistance corresponds to quantities such as commute times or escape probabilities \cite{doyle1984random, Tetali1991, Chandra1996}. In applied contexts, they have been used to model phenomena including dielectric breakdown \cite{PhysRevE.67.066610}, blood flow in vascular networks \cite{PhysRevE.99.012321}, the electrical properties of metallic nanowire meshes \cite{Wang2021}, and the geometry-dependent behavior of topological Josephson junctions \cite{Molenaar_2014}. Mathematically, the governing equations involve the discrete Laplace operator, which is structurally analogous to the tight-binding Hamiltonians widely used in solid-state physics. Therefore, when extended to include not only resistive but also capacitive and inductive elements, such networks can mimic the electronic band structure of crystals. For example, topolectrical circuits have attracted growing interest as accessible experimental platforms that reproduce key features of topological insulators \cite{Lee2018, PhysRevResearch.3.023056, 10.1063/5.0265293}. Remarkably, they make it possible to observe and investigate edge or surface states in simple, tabletop classical systems.

\begin{figure}[h!]
\centering
\includegraphics[width=0.8\columnwidth]{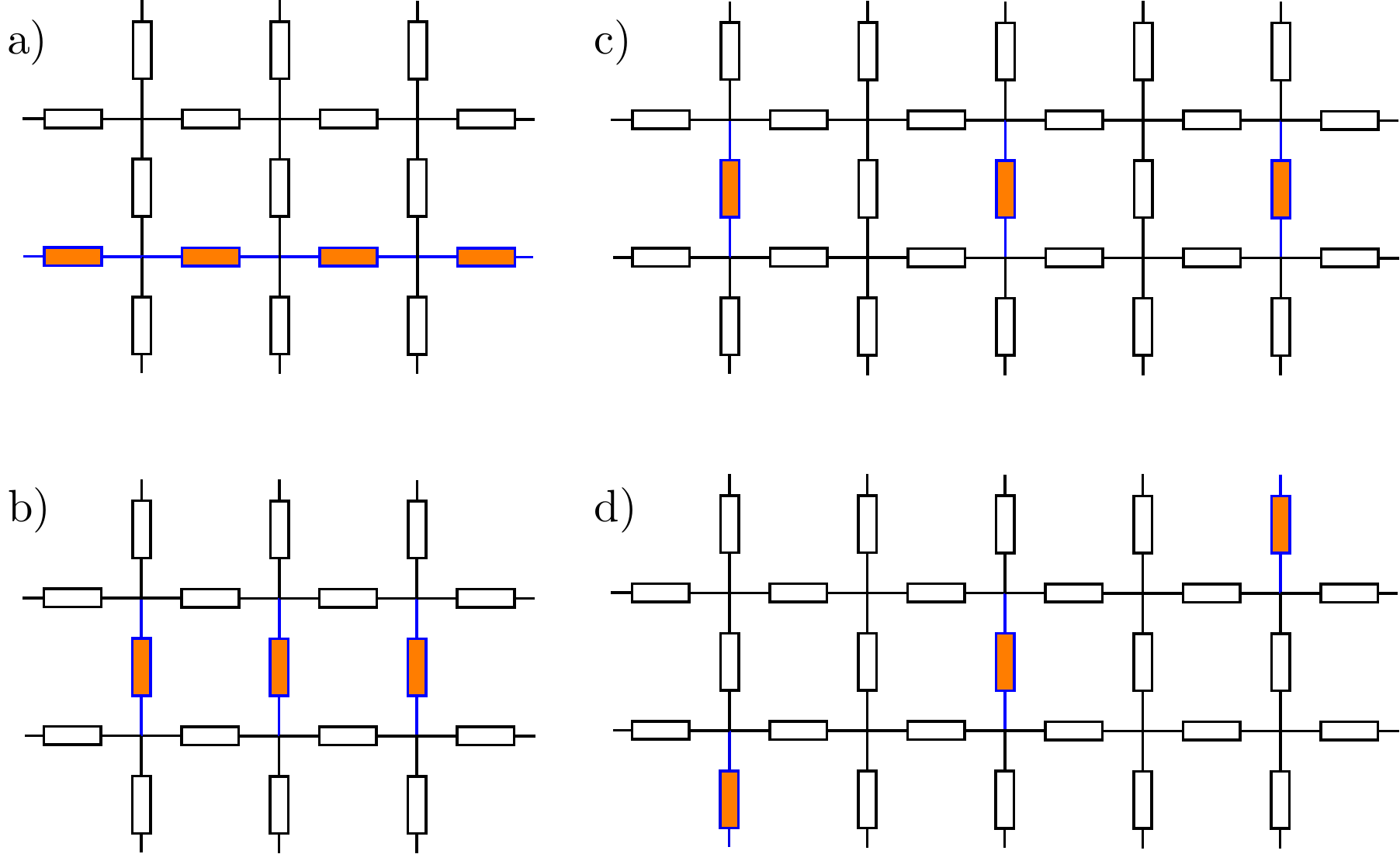}
\caption {\label{fig:general} \footnotesize{\textbf{Different types of line defects in an infinite resistor lattice.} In a perfect square lattice, the resistors located periodically along a given line are modified identically. Modified resistors are represented by filled rectangles.  For each of the configurations illustrated, this work shows how to compute the equivalent resistance between any two points.}}
\end{figure}

When a single resistor \cite{10.1119/1.1419104}, or a finite number of them \cite{cserti2025generaltheoryperturbationinfinite}, is removed or modified in a perfect infinite resistor network, the Green's function formalism can be extended using the Sherman–Morrison–Woodbury formula \cite{Numerical_Recipes_3rd_10.5555:book, guttman1946enlargement, Woodbury:cikk}, yielding solutions for the effective resistance in the presence of such defects. However, certain applications, such as two-dimensional topolectrical circuits, require evaluating the effective resistance in the vicinity of a one-dimensional boundary of the system.  In such cases, an infinite number of periodically arranged resistors are altered or removed, posing new analytical challenges. For examples, see figure~\ref{fig:general}.  In this work, we develop a general framework to address this class of problems by combining the reciprocal space representation of the Green’s function with the Sherman–Morrison identity \cite{sherman1950adjustment}. We demonstrate the method on the infinite square lattice, though it is readily applicable to other geometries and can be extended to networks involving complex impedances rather than purely resistive elements.

The structure of this paper is as follows. In Section 2, we review the well-established calculation of effective resistance on the infinite two-dimensional square lattice using the Green’s function formalism. 
In Sections 3 and 4, we then study two different configurations of one-dimensional horizontal line defects, where resistors on the horizontal (figure~\ref{fig:general}a) or vertical (figure~\ref{fig:general}b) edges are modified. The case of alternating modifications (figure~\ref{fig:general}c) is treated in Section 5. Finally, the most general scenario—featuring a line defect oriented at an arbitrary angle (figure~\ref{fig:general}d)—is addressed in Section 6. The paper concludes with a brief summary.

\section{Short review: perfect infinite square lattice}

In this chapter, we review the classical problem of calculating the effective resistance between two arbitrary nodes in an infinite square lattice composed of identical resistors of resistance $R$. The calculation is carried out using Green’s function techniques, following mainly reference~\cite{10.1119/1.1285881}. The primary purpose of this chapter is to establish the notational framework that will be used later on in this work.

The vertices of the square lattice are indexed by their Cartesian coordinates $(x,y)$. For simplicity, the lattice spacing is set to unity. For mathematical convenience, we first consider a finite lattice of size $N_{x(y)}$ in the $x$ ($y$) direction with periodic boundary conditions. After obtaining the relevant physical quantities, the limit $N_{x(y)}\to \infty$ will be taken to recover the case of an infinite square lattice.

We denote the electric potential at node $(x,y)$ by $V(x,y)$, and the externally injected (positive) or extracted (negative) current at that node by $I(x,y)$.
Adopting the bra–ket formalism familiar from quantum mechanics, which has also been used in previous studies of resistor networks, we associate a ket vector $|x,y\rangle$ to each node. These vectors form a complete orthonormal basis on an abstract Hilbert space.

Following Kirchhoff’s and Ohm’s laws, the governing equation for the node potentials can be expressed as
\begin{equation}
\hat{L}_0 |V\rangle = -R |I\rangle,
\end{equation}
where the potential and current vectors are given by
\begin{equation}
|V\rangle = \sum_{x = 1}^{N_x} \sum_{y = 1}^{N_y} V(x,y)\, |x,y\rangle, \qquad |I\rangle = \sum_{x = 1}^{N_x} \sum_{y = 1}^{N_y} I(x,y)\, |x,y\rangle.
\end{equation} Here $\hat{L}_0$ is the discrete Laplace operator of the resistor network and is given by
\begin{equation}
\begin{aligned}
\hat{L}_0 = \sum_{x = 1}^{N_x} \sum_{y = 1}^{N_y} \Big( & |x+1,y\rangle\langle x,y| + |x-1,y\rangle\langle x,y| + \\
& + |x,y+1\rangle\langle x,y| + |x,y-1\rangle\langle x,y| - 4\,|x,y\rangle\langle x,y| \Big),
\end{aligned}
\label{eq:laplacian}
\end{equation} where addition in the labels is defined modulo $N_{x(y)}$ due to the periodic boundary conditions.

Next, we introduce the lattice Green's function, which is defined as $\hat{G}_0\hat{L}_0=-1$. Using this function, the potential of the nodes is obtained by the following equation: $|V\rangle = R\hat{G}_0|I \rangle $ . If an electric current of magnitude $I$ enters at the point $(x_\textrm{in},y_\textrm{in})$ and exits at the point $(x_\textrm{out},y_\textrm{out})$, then the effective resistance between these two points is given by the expression:
\begin{equation}
\begin{aligned}
&R_0(x_\textrm{out},y_\textrm{out};x_\textrm{in},y_\textrm{in})=\frac{V(x_\textrm{in},y_\textrm{in})-V(x_\textrm{out},y_\textrm{out})}{I}\\&=R\left[ G_0(x_\textrm{in},y_\textrm{in}; x_\textrm{in},y_\textrm{in}) + G_0(x_\textrm{out},y_\textrm{out};x_\textrm{out},y_\textrm{out})-2 G_0(x_\textrm{in},y_\textrm{in}; x_\textrm{out},y_\textrm{out}) \right],
\end{aligned}
\label{eq_equivalent_resistance}
\end{equation}
where we use the notation $G_0(x_1, y_1; x_2,y_2)=\langle x_1,y_1 | \hat{G}_0 | x_2,y_2 \rangle$ for the real space matrix elements of the lattice Green's function.

The translational symmetry of the square lattice allows the problem to be treated more easily in reciprocal space; therefore, we introduce the Hilbert space vectors associated with the wave vectors $\boldsymbol{k}=(k_x,k_y)$:
\begin{equation}
|k_x,k_y\rangle =\frac{1}{\sqrt{N_x N_y}} \sum_{x = 1}^{N_x} \sum_{y = 1}^{N_y} \euler^{\iu k_xx} \euler^{\iu k_yy} |x,y\rangle .
\label{eq:plane_waves}
\end{equation} The wave vector is limited in the first Brillouin-zone and $k_{x(y)} = 2\pi n / N_{x(y)}$ with $n\in \mathbb Z$. In the following, we denote these respective sets by $\mathcal{B}_{x(y)}$, while the total Brillouin zone is given by their Cartesian product $\mathcal{B} = \mathcal{B}_x \times \mathcal{B}_y$.

In reciprocal space, both the lattice Laplacian and the Green's function are diagonal matrices. The latter one has the form 
\begin{equation}
\begin{aligned}
&\hat{G}_0 = \sum_{k_x \in \mathcal{B}_x} \sum_{k_y \in \mathcal{B}_y} G_0(k_x,k_y) \, |k_x,k_y\rangle \langle k_x,k_y|,\\
&G_0(k_x,k_y)=\frac{1}{4-2\cos{k_x}-2\cos{k_y}}.
\end{aligned}
\label{eq:green_operator}
\end{equation} One can note that the coefficients $G_0(k_x,k_y)$ are singular for $k_x = k_y = 0$ which seems to imply that $\hat{G}_0$ does not exist. Strictly speaking, this is true as the Laplacian $\hat{L}_0$ defined in equation~\eqref{eq:laplacian} is indeed a non-invertible operator. However, as was shown in previous works, including reference \cite{cserti2025generaltheoryperturbationinfinite}, a minor modification can make the Laplacian invertible while keeping it physically equivalent to equation~\eqref{eq:laplacian}. Since none of our results would be modified by such a redefinition, we continue to use equation~\eqref{eq:green_operator} throughout this work.

Using the explicit form of plane wave states from equation~\eqref{eq:plane_waves}, we can express the real space coefficients of the Green's function:
\begin{equation}
G_0(x_1, y_1; x_2,y_2) = \frac{1}{N_x N_y} \sum_{k_x \in \mathcal{B}_x} \sum_{k_y \in \mathcal{B}_y} G_0(k_x,k_y) \euler^{\iu k_x(x_2-x_1) + \iu k_y(y_2-y_1)}.
\label{eq:Fourier}
\end{equation}

To study perturbations of an infinite resistor network, one has to take the limit $N_{x(y)} \to \infty$. In this limit, the summations appearing in the Green's function become integrals over the Brillouin zone:
\begin{equation}
    \frac{1}{N_x} \sum_{k_x \in \mathcal{B}_x} \longrightarrow \frac{1}{2\pi} \int_{-\pi}^\pi \mathrm{d}k_x , \qquad \frac{1}{N_y} \sum_{k_y \in \mathcal{B}_y} \longrightarrow \frac{1}{2\pi} \int_{-\pi}^\pi \mathrm{d}k_y .
    \label{eq:thermo_lim}
\end{equation} Making this substitution in equation~\eqref{eq:Fourier}, we find that
\begin{equation}
G_0(x_1, y_1; x_2,y_2)=\frac{1}{(2\pi)^2}\int_{-\pi}^\pi dk_x\int_{-\pi}^\pi dk_y \, G_0(k_x,k_y) \euler^{\iu k_x(x_2-x_1) + \iu k_y(y_2-y_1)}.
\end{equation}
It has been demonstrated in earlier works that, for the square lattice, one of the integrals can be carried out analytically,
\begin{equation}
G_0(k_x,y_1,y_2)\equiv \frac{1}{2\pi} \int_{-\pi}^\pi dk_y \, G_0(k_x,k_y) \euler^{\iu k_y(y_2-y_1)} =
\frac{1}{2}\frac{\euler^{-|y_2-y_1|s}}{\sinh{s}},
\label{eq_perfect_green}
\end{equation}
where $\cosh{s}=2-\cos{k_x}$ \cite{10.1119/1.17696, 10.1119/1.19311}.
Finally, the effective resistance between the points $(x_\textrm{in},y_\textrm{in})$ and  $(x_\textrm{out},y_\textrm{out})$ can be expressed by an integral formula, which admits an analytical solution for arbitrary points:
\begin{equation}
R_0(x_\textrm{out},y_\textrm{out};x_\textrm{in},y_\textrm{in})= R  \int_{-\pi}^\pi \frac{dk_x }{2\pi}\frac{1-\euler^{-|y_\textrm{out}-y_\textrm{in}|s}\cos{\left[k_x (x_\textrm{out}-x_\textrm{in}) \right]}}{\sinh{s}}.
\end{equation}
This integral has been the subject of extensive analysis in prior studies, where its asymptotic behavior was characterized and recurrence relations were derived to facilitate its computation \cite{10.1119/1.1285881}. Due to the translational symmetry of the lattice, the integral depends only on the relative position of the input and output points, $R_0(x_\textrm{out}-x_\textrm{in},y_\textrm{out}-y_\textrm{in}) \equiv R_0(x_\textrm{out},y_\textrm{out};x_\textrm{in},y_\textrm{in})$. Some specific values of the effective resistance are given in table~\ref{perfect:table} for $\Delta x = x_\textrm{out} - x_\textrm{in}$ and $\Delta y = y_\textrm{out} - y_\textrm{in}$.

\begin{table}[hbt]
	\caption{\label{perfect:table}
			Resistance $R_0(\Delta x, \Delta y)$ of a perfect infinite square lattice in units of $R$.}
	\begin{indented}
		\item[]\begin{tabular}{@{}c|c c c c}
			\br
			$\Delta x$ / $\Delta y$ & $0$ & $1$ & $2$ & $3$ \\
			\mr
            $0$ & $0$ & $\frac12$ & $2 - \frac{4}{\pi}$ & $\frac{17}{2} - \frac{24}{\pi}$ \\[2ex]
            $1$ & $\frac{1}{2}$ & $\frac{2}{\pi}$ & $\frac{4}{\pi} - \frac12$ & $\frac{46}{3\pi} - 4$ \\[2ex]
			$2$ & $2-\frac{4}{\pi}$ & $\frac{4}{\pi} - \frac12$ & $\frac{8}{3\pi}$ & $\frac12 + \frac{4}{3\pi}$   \\[2ex]
            $3$ & $\frac{17}{2} - \frac{24}{\pi}$ & $\frac{46}{3\pi} - 4$ & $\frac12 + \frac{4}{3\pi}$ & $\frac{46}{15\pi}$ \\[2ex]
			\br
		\end{tabular}
	\end{indented}
\end{table}

\section{Parallel line defect}
\label{chapter:parallel}

\begin{figure}[b]
\centering
\includegraphics[width=0.5\columnwidth]{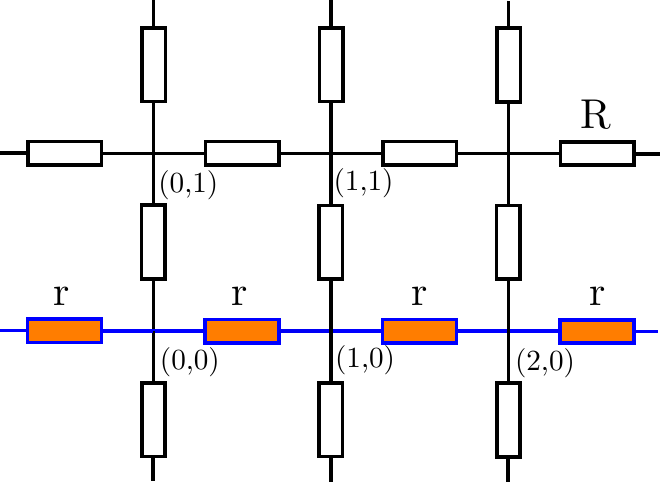}
\caption {\label{fig:vertical} \footnotesize{\textbf{Parallel line defect in an infinite resistor lattice (modification on horizontal edges).} A subset of the resistors, originally of resistance $R$ on the horizontal edges, have been replaced by resistors $r$ along a horizontal line, as indicated. Several node coordinates $(x,y)$ are also labeled. The origin is chosen as one of the nodes along the line defect.}}
\end{figure}

In this chapter, we consider an infinite square lattice in which all resistors of resistance $R$ along a horizontal line are replaced by resistors of resistance $r$, as illustrated in figure~\ref{fig:vertical}. Now and later in this paper, we always choose the origin as an endpoint of one of the perturbed links. Our objective is to derive an analytical expression for the equivalent resistance between two arbitrary points in such a modified lattice.

As a first step, based on \cite{10.1119/1.1419104}, we consider how to account for the replacement of a single resistor of resistance $R$ between neighboring points $(x_1,y_1)$ and $(x_2,y_2)$ with a resistor of resistance $r$. The original network has the same potential distribution as the modified network if an external current 
\begin{equation}
\delta I = \frac{r-R}{rR}\left[ V(x_1,y_1) - V(x_2,y_2) \right]
\end{equation}
enters at the point $(x_1,y_1)$ and exits at the point $(x_2,y_2)$ in the original network. For later purposes, we introduce the parameter $g = 1 - R/r$ to characterize the strength of the perturbation, where $g\in (-\infty,1]$. If the resistor is completely removed from the network, $g=1$; if it is replaced by a short circuit, then $g\rightarrow-\infty$. The additional current contribution can be expressed as follows:
\begin{equation} \label{eq:curent_pert}
 \delta I  |x_1,y_1\rangle -\delta I |x_2,y_2\rangle =\frac{g}{R}   (|x_1,y_1\rangle-|x_2,y_2\rangle)(\langle x_1,y_1|-\langle x_2,y_2|) |V\rangle.
\end{equation}

In the case of the line defect, an infinite number of resistors must be replaced. Consequently, additional incoming and outgoing currents must be taken into account at each node of the form $(x,0)$. These perturbations can be captured by summing equation~(\ref{eq:curent_pert}) over the relevant links,
\begin{equation}
\begin{aligned}
\hat{L}_1|V\rangle &\equiv g \sum_{x = 1}^{N_x} (|x,0\rangle-|x+1,0\rangle)(\langle x,0|-\langle x+1,0|) |V\rangle \\
&= g \sum_{x = 1}^{N_x} \left(2|x,0\rangle\langle x,0| - |x,0\rangle\langle x+1,0|-|x+1,0\rangle\langle x,0|\right)|V\rangle.
\end{aligned}
\label{eq_perturbation_parallel}
\end{equation}
Thus, in the perturbed network, the  relationship between the node potentials and external currents can be written as follows:
\begin{equation}
\left(\hat{L}_0+\hat{L}_1\right)|V\rangle = -R |I\rangle.
\end{equation}

The perturbed lattice has translational invariance along the $x$-direction; therefore, we perform a Fourier transformation only in the $x$-direction and use the corresponding wave vector $k_x$,
\begin{equation} \label{eq:basis}
|k_x,y\rangle = |k_x\rangle \otimes |y\rangle =\frac{1}{\sqrt{N_x}} \sum_{x = 1}^{N_x} \euler^{\iu k_xx} |x,y\rangle .
\end{equation}
In this basis, the perturbation $\hat{L}_1$ is represented by a block-diagonal matrix,
\begin{equation}
    \hat{L}_1 = 2g \sum_{k_x\in\mathcal{B}_x} \left(1 - \cos{k_x}\right) |k_x,y = 0\rangle \langle k_x, y = 0| \equiv \sum_{k_x\in\mathcal{B}_x} |k_x\rangle\langle k_x| \otimes \hat{L}_1(k_x) ,
\label{eq_pertubation_blockform}
\end{equation} where the respective blocks are given as
\begin{equation}
    \hat{L}_1(k_x) = \left(1 - \cos{k_x}\right) |0\rangle \langle 0| .
\end{equation} In this formula, we used the shorthand notation $|0\rangle$ for the vector $|y = 0\rangle$. We shall keep this from now on for simplicity.
Similarly, we can give an implicit definition for the block operators of the Laplacian in equation~\eqref{eq:laplacian} by
\begin{equation}
    \hat{L}_0 = \sum_{k_x\in\mathcal{B}_x} |k_x\rangle\langle k_x| \otimes \hat{L}_0(k_x) .
\end{equation} The explicit definiton of takes the following form:
\begin{equation}
    \hat{L}_0(k_x) = \sum_{k_y\in\mathcal{B}_y} \left(2\cos k_x + 2\cos k_y - 4\right) \, |k_y\rangle \langle k_y| .
\end{equation}

Owing to the block-diagonal structure, taking the inverse consists of inverting each block separately. For the Green's function of the perfect lattice, this allows us to introduce the operators $\hat{G}_0(k_x) = - \hat{L}_0^{-1}(k_x)$. For the Green's function of the perturbed lattice, this leads to $\hat{G}(k_x)=-[\hat L_0(k_x)+\hat L_1(k_x)]^{-1}$. The latter can be readily computed using the Sherman--Morrison formula \cite{sherman1950adjustment}. This identity states that for arbitrary vectors $|u\rangle$, $|v\rangle$, and an invertible operator $\hat{A}$:
\begin{equation}
    \left(\hat{A} + |u\rangle \langle v|\right)^{-1} = \hat{A}^{-1} - \frac{\hat{A}^{-1}|u\rangle \langle v|\hat{A}^{-1}}{1 + \langle v|\hat{A}^{-1}|u\rangle} .
    \label{eq:Sherman-Morrison}
\end{equation} The inverse exists if and only if $1 + \langle v|\hat{A}^{-1}|u\rangle \neq 0$. Identifying $\hat{A}$ with $\hat{L}_0(k_x)$, $|u\rangle$ with $(1 - \cos k_x)|0\rangle$, and $|v\rangle$ with $|0\rangle$, we find the following:
\begin{equation}
    \hat{G}(k_x) = \hat{G}_0(k_x) + \frac{2g \left(1 - \cos{k_x}\right) \hat{G}_0(k_x) | 0\rangle\langle 0| \hat{G}_0(k_x)}{1 - 2g \left(1 - \cos{k_x}\right) \langle 0| \hat{G}_0(k_x) | 0\rangle} .
    \label{eq:sm_horizontal_1}
\end{equation} Note that the vector $|y = 0\rangle$ appears in the result since the perturbation is located along the $y = 0$ line. To proceed with the calculation, we can make use of the matrix elements of the block operators of the Green's function:
\begin{equation} \label{eq:Green_basis}
\hat{G}(k_x) = \sum _{y_1 = 1}^{N_y} \sum _{y_2 = 1}^{N_y} G(k_x,y_1,y_2)|y_1\rangle \langle y_2|.
\end{equation} In this representation, equation~\eqref{eq:sm_horizontal_1} takes the following form:
\begin{equation}
    G(k_x,y_1,y_2) = G_0(k_x,y_1,y_2) + \frac{2g \left(1 - \cos{k_x}\right) G_0(k_x,y_1,0) G_0(k_x,0,y_2)}{1 - 2g \left(1 - \cos{k_x}\right) G_0(k_x,0,0)} .
    \label{eq:sm_horizontal_2}
\end{equation}

Taking the limit $N_{x(y)} \to \infty$ according to equation~\eqref{eq:thermo_lim}, and using equation~\eqref{eq_perfect_green} for the matrix elements in the case of the perfect lattice, the denominator of equation~\eqref{eq:sm_horizontal_2} can be expressed as:
\begin{equation}
    1 - 2g \left(1 - \cos{k_x}\right) G_0(k_x,0,0) = 1 - g\cdot\frac{\sinh{s}}{\cosh{s} + 1} .
\end{equation} The numerator is obtained similarly:
\begin{equation}
    2g \left(1 - \cos{k_x}\right) G_0(k_x,y_1,0) G_0(k_x,0,y_2) = \frac{g}{2}\cdot \frac{\euler^{-(|y_1| + |y_2|)s}}{\cosh{s} + 1} .
\end{equation} Using both of these results, we find that the matrix elements of the Green's function are
\begin{equation}
    G(k_x,y_1,y_2) = G_0(k_x,y_1,y_2) + \frac{g}{2}\cdot \frac{\euler^{-(|y_1| + |y_2|)s}}{\cosh{s} + 1 - g\sinh{s}} .
\end{equation}

Finally, we can transform this back into real space by an inverse Fourier transform with respect to the $k_x$ variable. This leads to
\begin{equation} \label{eq:green_function_paralel}
    G(x_1,y_1;x_2,y_2) = G_0(x_1,y_1;x_2,y_2) + \frac{g}{4\pi} \int_{-\pi}^\pi \mathrm{d}k_x\, \frac{\euler^{-(|y_1| + |y_2|)s} \cos\left[k_x(x_1-x_2)\right]}{\cosh{s} + 1 - g\sinh{s}} .
\end{equation} By substituting this expression into equation~\eqref{eq_equivalent_resistance}, we can determine the effective resistance between points $(x_\mathrm{in},y_\mathrm{in})$ and $(x_\mathrm{out},y_\mathrm{out})$ within the perturbed lattice:
\begin{equation}
    \begin{aligned}
        R(x_\mathrm{out},y_\mathrm{out}&;x_\mathrm{in},y_\mathrm{in}) = R_0(x_\mathrm{out},y_\mathrm{out};x_\mathrm{in},y_\mathrm{in}) \\ &+ \frac{gR}{4\pi} \int_{-\pi}^\pi \mathrm{d}k_x\, \frac{\euler^{-2|y_\mathrm{out}|s} + \euler^{-2|y_\mathrm{in}|s} - 2 \euler^{-(|y_\mathrm{out}| + |y_\mathrm{in}|)s} \cos\left[k_x(x_\mathrm{out}-x_\mathrm{in})\right]}{\cosh{s} + 1 - g\sinh{s}} .
    \end{aligned}
    \label{eq_resistance_horizontal}
\end{equation}

The integral expression obtained for the effective resistance is the main result of this chapter. It provides a numerically efficient method for computing the effective resistance in the presence of parallel line defects. While performing the integral analytically is challenging in general, the following subsections demonstrate that for specific values of $g$, or between nodes located near the line defect, the effective resistance can be evaluated.

Furthermore, due to the symmetries of the system, the effective resistance satisfies several important properties. First, because the lattice is invariant under horizontal translations and reflections about vertical axes, the resistance between two nodes depends only on the horizontal distance $|x_\textrm{out}-x_\textrm{in}|$ between them. In addition, the system is symmetric under reflection across the line defect. This implies that $R(x_\mathrm{out},y_\mathrm{out};x_\mathrm{in},y_\mathrm{in})=R(x_\mathrm{out},-y_\mathrm{out};x_\mathrm{in},-y_\mathrm{in})$. Specifically, if the input and output nodes are placed symmetrically with respect to the defect line, then all the nodes along the defect become equipotential. Therefore, the effective resistance is independent of the value of $g$. Taking $g=0$, we find that the effective resistance corresponds to the well-known values for the unperturbed system,
\begin{equation}
R(x_\mathrm{in},-y_\mathrm{in};x_\mathrm{in},y_\mathrm{in})=R_0(0,2y_\textrm{in}).
\end{equation}

\subsection{Analytical results for special $g$ values}

As it was pointed out before, evaluating the integral of equation~\eqref{eq_resistance_horizontal} in general is a challenging task, and the second term is probably not always reducible to the equivalent resistance $R_0$ of the perfect lattice. However, there are a couple of special cases, which can be easily computed: (i) $g = 0$, the trivial case; (ii) $g \to -\infty$, where the resistors along the defect line are short-circuited; (iii) $g = 1$, where the resistors along the defect line are short-circuited; and (iv) $g = -1$, where the line defect consists of resistors with resistance $R/2$. 

First, let us take $g = 0$ (i.e., $r = R$) when none of the resistances are modified. In this case, the second term vanishes completely, recovering the expected result: $R(x_\mathrm{out},y_\mathrm{out};x_\mathrm{in},y_\mathrm{in}) = R_0(x_\mathrm{out}-x_\mathrm{in}, y_\mathrm{out}-y_\mathrm{in})$.
    
Now take $g \to -\infty$ (i.e., $r = 0$) when all resistors on a horizontal line are short-circuited. In this limit, the last term in the denominator of equation~\eqref{eq:green_function_paralel} is asymptotically dominant, leading to
\begin{equation} \label{eq:Green_parallel}
G(x_1,y_1;x_2,y_2)=\begin{cases}
0 &\text{if $\textrm{sgn}({y_1})\neq \textrm{sgn}({y_2})$},\\
G_0(x_1,y_1;x_2,y_2)-G_0(x_1,-y_1;x_2,y_2)&\text{if $\textrm{sgn}({y_1})=\textrm{sgn}({y_2})$},
\end{cases}
\end{equation}
where sgn is the signum function. The potential of a short-circuited horizontal line defect is always zero. Therefore, if a current $I$ is injected into the lattice in one region (say, above the defect line), it does not propagate to the other region (below the defect line). Consequently, the potential in the other region is identically zero, which explains why the Green's function evaluates to zero in the first case of equation~(\ref{eq:Green_parallel}). 

In the second case, where we examine the potential on the same side of the defect where the current is injected, the result can be understood using a symmetry argument. According to the previous paragraph, the node potentials on this side remain unchanged if we extract a current $I$ in the mirror image of the injection point with respect to the line defect. Then, due to the symmetry of the system, the potential along the defect remains zero for arbitrary values of the parameter $g$. Furthermore, the current and voltage distributions in the upper region remain identical to those in the original problem, as guaranteed by the uniqueness theorem for Poisson's equation. In the special case $g=0$, the Green's function for the perturbed lattice can be expressed as the difference between the Green's functions of the perfect square lattice.

By substituting the Green's function into equation~(\ref{eq_equivalent_resistance}), the resistance of the perturbed lattice can be expressed in terms of the resistance of the perfect square lattice:
\begin{equation}
    \begin{aligned}
        R(x_\mathrm{out},y_\mathrm{out};x_\mathrm{in},y_\mathrm{in}) &= R_0(x_\mathrm{out}-x_\mathrm{in},y_\mathrm{out}-y_\mathrm{in}) - R_0(x_\mathrm{out}-x_\mathrm{in},|y_\mathrm{out}|+|y_\mathrm{in}|)\\ &+ \frac12 R_0(0,2y_\mathrm{in}) + \frac12 R_0(0,2y_\mathrm{out}).
    \end{aligned}
\end{equation}
Using this expression and table~\ref{perfect:table}, we can obtain exact values for certain special cases,
\begin{equation}
    \begin{aligned}
    R(1,0;0,0) &= 0,\\
    R(0,1;0,0) &= \frac{1}{2} R_0(0,2)=1-\frac{2}{\pi},\\
     R(1,1;0,1) &= R_0(1,0)-R_0(1,2)+R_0(0,2) = 3 - \frac{8}{\pi}.\\
   \end{aligned}
   \label{eq:res_par_g=-inf}
\end{equation}
For simplicity, the unit $R$ is omitted.

Now take $g = 1$ (i.e., $r \to \infty$) when all resistors on a horizontal line are removed from the lattice. In this limit, the denominator  of equation~(\ref{eq:green_function_paralel}) reduces to $1 + \euler^{-s}$. Using the hyperbolic identity
\begin{equation}
    \frac{1}{1 + \euler^{-s}} = \frac{1}{\sinh{s}}\frac{\euler^s - 1}{2},
\end{equation} we can transform the Green's function as follows:
\begin{equation}
    \begin{aligned}
G(x_1,y_1;x_2,y_2) &= G_0(x_1,y_1;x_2,y_2) \\ &+ \frac{1}{2} \left[ G_0(x_1,0;x_2,|y_2|+|y_1|-1)-G_0(x_1,0;x_2,|y_2|+|y_1|)\right].
  \end{aligned}
\end{equation}
We should note that this transformation is valid only when $y_1$ and $y_2$ are both non-zero.
By substituting the Green's function into equation~(\ref{eq_equivalent_resistance}), the resistance of the perturbed lattice can be expressed in terms of the resistance of the perfect square lattice.
\begin{equation}
\begin{aligned}    R(x_\mathrm{out},y_\mathrm{out};x_\mathrm{in},y_\mathrm{in}) &= R_0(x_\mathrm{out}-x_\mathrm{in},y_\mathrm{out}-y_\mathrm{in}) \\ &+ \frac12 R_0(x_\mathrm{out}-x_\mathrm{in},|y_\mathrm{in}| + |y_\mathrm{out}| - 1) \\ &- \frac12 R_0(x_\mathrm{out}-x_\mathrm{in},|y_\mathrm{in}| + |y_\mathrm{out}|) - \frac14 R_0(0,2|y_\mathrm{out}| - 1) \\ &- \frac14 R_0(0,2|y_\mathrm{in}| - 1) + \frac14 R_0(0,2y_\mathrm{out}) + \frac14 R_0(0,2y_\mathrm{in}).
\end{aligned}
\label{eq:resistance_g1}
\end{equation}
This expression is valid only when both $y_\textrm{in}$ and $y_\textrm{out}$ are non-zero. Through a more careful analysis of equation~\eqref{eq_resistance_horizontal}, it can be shown that if either $y_\textrm{in}$ or $y_\textrm{out}$ is zero, then an additional term of $R/2$ must be added to the expression of the equivalent resistance given in equation~(\ref{eq:resistance_g1}). Furthermore, if both $y_\textrm{in}$ and $y_\textrm{out}$ are zero, a correction term of $R$ must be added. In the following, we provide analytical expressions for several effective resistances in units of $R$ using table~\ref{perfect:table}:
\begin{equation}
    \begin{aligned}
    R(1,0;0,0) &= 1+\frac{1}{2}R_0(1,1)=1+\frac{1}{\pi},\\
    R(0,1;0,0) &= \frac{1}{2}+\frac{1}{4} R_0(0,2)=1-\frac{1}{\pi},\\
     R(1,1;0,1) &= \frac{1}{2}\left[R_0(0,1)+R_0(1,1)+R_0(0,2)-R_0(1,2)\right]=\frac32-\frac{3}{\pi}.\\
   \end{aligned}
   \label{eq:res_par_g=1}
\end{equation}

Finally, the last case, which can be readily reduced to the perfect square lattice scenario, arises when all resistors are replaced by $r = R/2$, corresponding to $g=-1$. In this limit, the denominator in equation~(\ref{eq:green_function_paralel}) reduces to $1 + \euler^{s}$. By a similar derivation as in the $g = 1$ case, but using a different identity
\begin{equation}
    \frac{1}{1 + \euler^{s}} = \frac{1}{\sinh{s}}\frac{1 - \euler^{-s}}{2}
\end{equation} this time, we arrive at the following result for the Green's function
\begin{equation}
    \begin{aligned}
G(x_1,y_1;x_2,y_2) &= G_0(x_1,y_1;x_2,y_2) \\ &+ \frac{1}{2} \left[G_0(x_1,0;x_2,|y_2|+|y_1|+1) - G_0(x_1,0;x_2,|y_2|+|y_1|)\right].
  \end{aligned}
\end{equation}
and for the
equivalent resistance:
\begin{equation}
    \begin{aligned}R(x_\mathrm{out},y_\mathrm{out};x_\mathrm{in},y_\mathrm{in}) &= R_0(x_\mathrm{out}-x_\mathrm{in},y_\mathrm{out}-y_\mathrm{in}) \\ &+ \frac12 R_0(x_\mathrm{out}-x_\mathrm{in},|y_\mathrm{in}| + |y_\mathrm{out}| + 1) \\ &-\frac12 R_0(x_\mathrm{out}-x_\mathrm{in},|y_\mathrm{in}| + |y_\mathrm{out}|) - \frac14 R_0(0,2|y_\mathrm{out}| + 1) \\ &- \frac14 R_0(0,2|y_\mathrm{in}| + 1) + \frac14 R_0(0,2|y_\mathrm{out}|)+ \frac14 R_0(0,2|y_\mathrm{in}|)  .
    \end{aligned}
    \label{eq_resistor_paralel_min1}
\end{equation}
Several effective resistances are evaluated analytically in the following, using table~\ref{perfect:table}:
\begin{equation}
    \begin{aligned}
    R(1,0;0,0) &= \frac{1}{2}R_0(1,1)=\frac{1}{\pi},\\
    R(0,1;0,0) &= \frac{1}{4}\left[R_0(0,1)+3R_0(0,2)-R_0(0,3)\right] = \frac{3}{\pi} - \frac{1}{2} ,\\
     R(1,1;0,1) &= \frac{1}{2}\left[2R_0(0,1)+R_0(1,3)-R_0(1,2)-R_0(0,3)+R_0(0,2)\right] = \frac{47}{3\pi} - \frac92.\\
   \end{aligned}
   \label{eq:res_par_g=-1}
\end{equation}

\subsection{Analytical results for special node coordinates}

For small values of $x_\textrm{out}-x_\textrm{in}$, $y_\textrm{in}$, and $y_\textrm{out}$, the integral given in equation~(\ref{eq_resistance_horizontal}) can be analytically evaluated for an arbitrary value of $g$. For example,
\begin{equation}
    \begin{aligned}
        R(1,0;0,0) &=  \frac{1}{1-g^2}\left[\frac{1+2g}{2}-F^\parallel(g)\right],\\ 
         R(0,1;0,0) &=   \frac{1}{1+g^2}\left[\frac{2g(1+g)(\pi-2)+\pi}{2\pi}+F^\parallel(g)\right]
        ,
    \end{aligned}
    \label{eq:resistance_par_analytical}
\end{equation}
where
\begin{equation} 
F^\parallel(g)=\begin{cases}
\frac{2g}{\pi\sqrt{g^2-2}}\textrm{arccosh}\left(\frac{-g}{\sqrt{2}}\right) &\text{if $g<-\sqrt{2}$},\\
\frac{2g}{\pi\sqrt{2-g^2}}\left[\frac{\pi}{2}+\textrm{arcsin}\left(\frac{g}{\sqrt{2}}\right)\right] &\text{if $g\geq -\sqrt{2}$}.
\end{cases}
\end{equation}

\begin{figure}[h!]
\centering
\includegraphics[width=0.6\columnwidth]{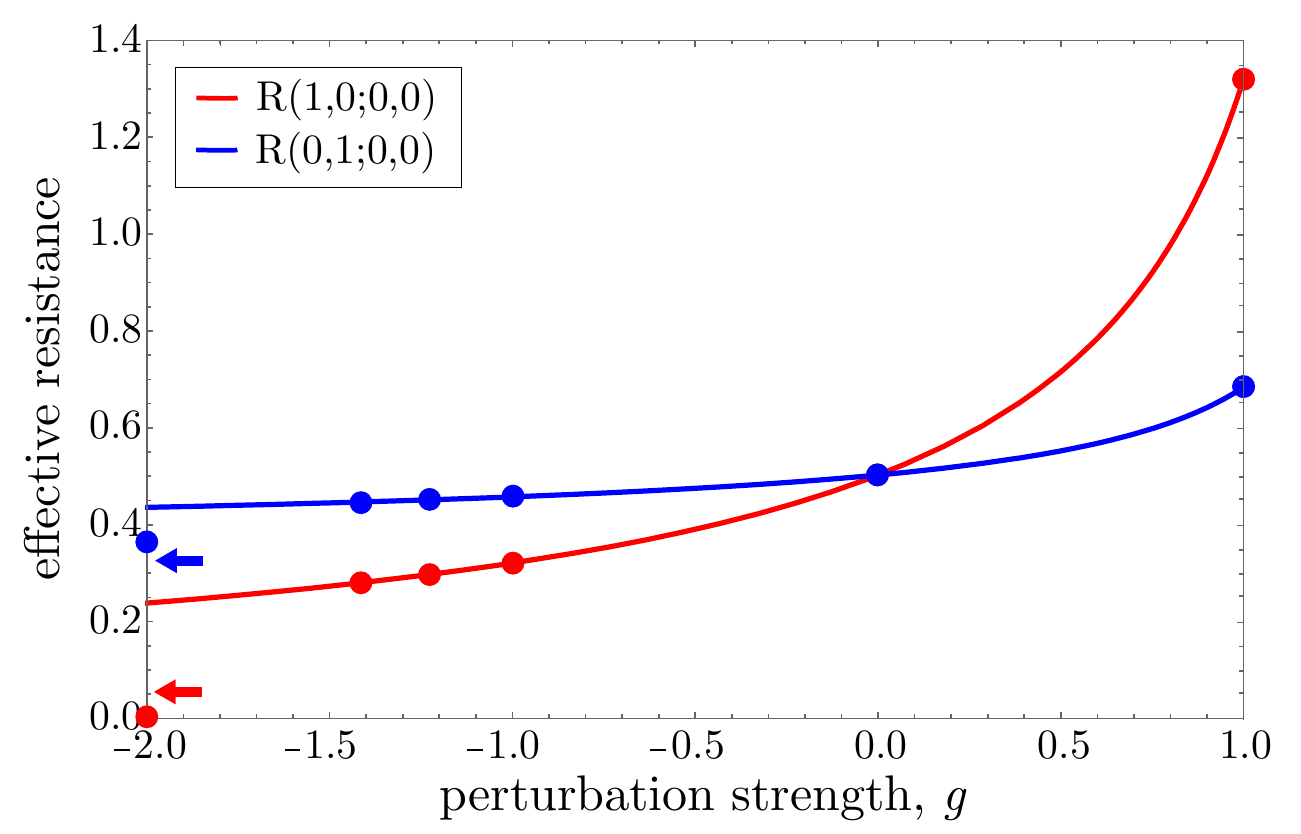}
\caption {\label{fig:parallel_result1} \footnotesize{\textbf{Effective resistances in the presence of a parallel line defect in an infinite resistor network} The red curve shows the dependence of the effective resistance on $g$ between the points $(0,0)$ and $(1,0)$, while the blue curve corresponds to the resistance between $(0,0)$ and $(0,1)$. The values listed in table \ref{R0:table} are indicated by discrete markers. On the left edge of the plot, the limiting values, $g\rightarrow-\infty$, of the effective resistance are also marked.}}
\end{figure}

In table \ref{R0:table}, we present the values of the effective resistance given by equation~\eqref{eq:resistance_par_analytical} for several specific values of $g$, in agreement with the intermediate results derived in equations~\eqref{eq:res_par_g=-inf}, \eqref{eq:res_par_g=1}, \eqref{eq:res_par_g=-1}. These selected points and the dependence of the effective resistance on the parameter $g$ are plotted in figure~\ref{fig:parallel_result1}.


\begin{table}[hbt]
	\caption{\label{R0:table}
			Resistance of an infinite square lattice in the presence of a parallel line defect in units of $R$.}
	\begin{indented}
		\item[]\begin{tabular}{@{}c|c|c}
			\br
			$g$ & $R(1,0;0,0)$ & $R(0,1;0,0)$ \\
			\mr
            $-\infty$ & $0$ & $1-\frac{2}{\pi}$ \\[2ex]
            $-\sqrt{2}$ & $\sqrt{2}-\frac{1}{2}-\frac{2}{\pi }$ & $\frac{7}{2}+2\sqrt{2}-\frac{10+6\sqrt{2}}{\pi }$ \\[2ex]
			$-\sqrt{3/2}$ & $\sqrt{6} - 1 - \frac{2}{\sqrt{3}}$ & $8+3\sqrt{6}-4\sqrt{2}-\frac{10}{\sqrt{3}}-\frac{6+2\sqrt{6}}{\pi}$   \\[2ex]
            $-1$ & $\frac{1}{\pi}$ & $\frac{3}{\pi} - \frac12$ \\[2ex]
            $0$ & $\frac12$ & $\frac12$ \\[2ex]
            $1$ & $1 + \frac{1}{\pi}$ & $1 - \frac{1}{\pi}$ \\[2ex]
			\br
		\end{tabular}
	\end{indented}
\end{table}

\subsection{Numerical results for the effective resistances}

\begin{figure}[h!]
\centering
\includegraphics[width=0.6\columnwidth]{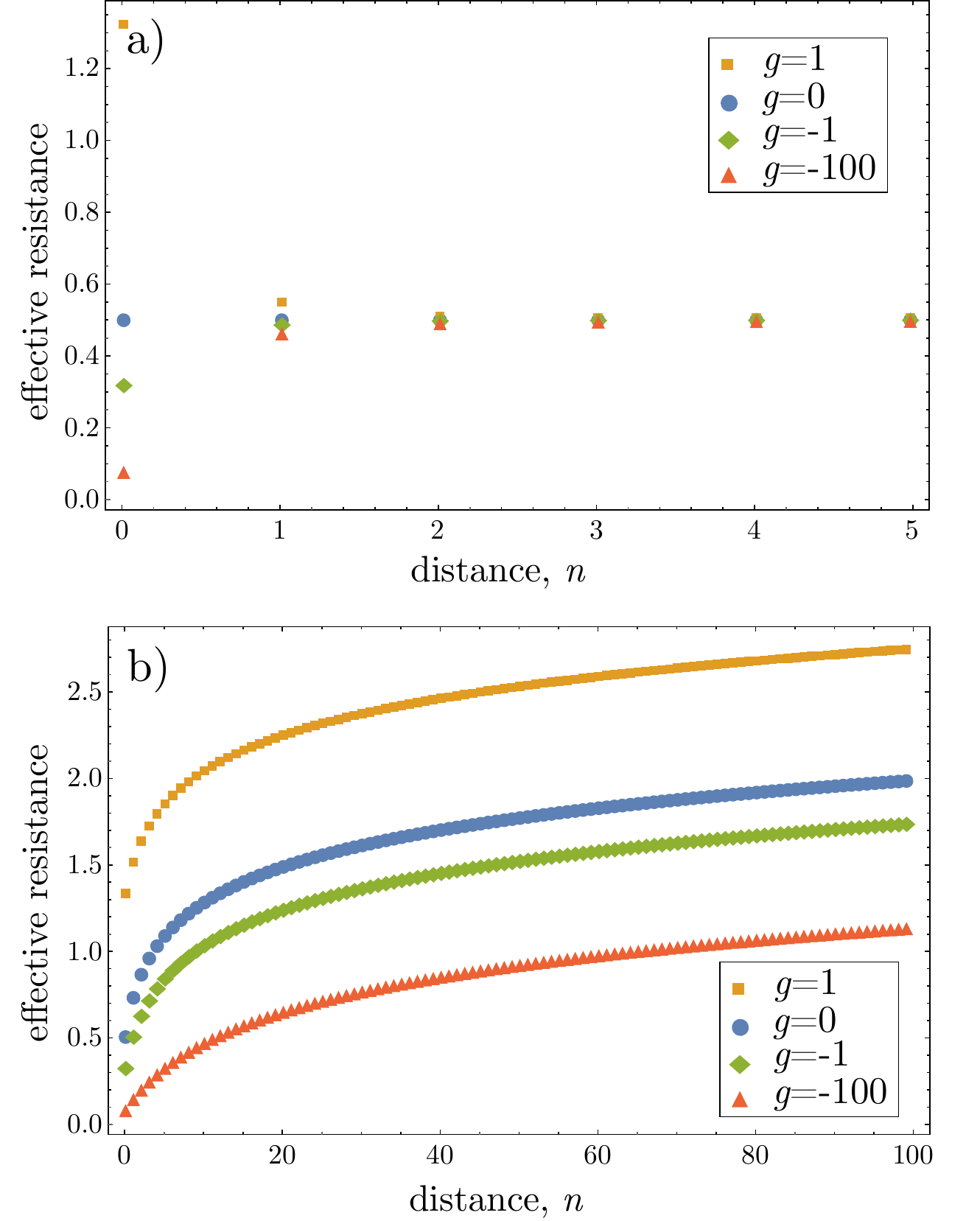}
\caption {\label{fig:parallel_figure4} \footnotesize{\textbf{Effective resistances in the presence of a parallel line defect in an infinite resistor network} a) The resistance $R(0,n;1,n)$, b) the resistance $R(0,0;n,0)$ in units of $R$ for different strength of the pertubation $g$.}}
\end{figure}

 As we saw in the previous subsections, evaluating the integral expression for the effective resistance is analytically challenging. However, since the integrand is nonsingular, the resistance values can be computed efficiently using numerical methods. To illustrate this, we carried out numerical calculations of the effective resistance in two specific geometric configurations. In figure~\ref{fig:parallel_figure4}a, we consider the resistance between two neighboring nodes, and study how this value changes with vertical distance from the defect line, i.e., $R(0,n;1,n)$. As the results show, already a few steps away from the defect, regardless of the strength of the perturbation $g$, the effective resistance quickly approaches the expected value of $R/2$. In figure~\ref{fig:parallel_figure4}b, we examined the resistance between nodes lying along the defect line, as a function of their horizontal separation, i.e., $R(0,0;n,0)$. In this case, the asymptotic behavior depends on the strength of the perturbation $g$, but in all cases, they seem to obey a similar dependence at large distances. For $g = 0$, it can be proven that the $n\to\infty$ asymptotic form is logarithmic \cite{10.1119/1.1285881, 10.1119/1.17696}, which seems to be inherited by the $g \neq 0$ cases as well.

\section{Perpendicular line defect}

\begin{figure}[h]
\centering
\includegraphics[width=0.5\columnwidth]{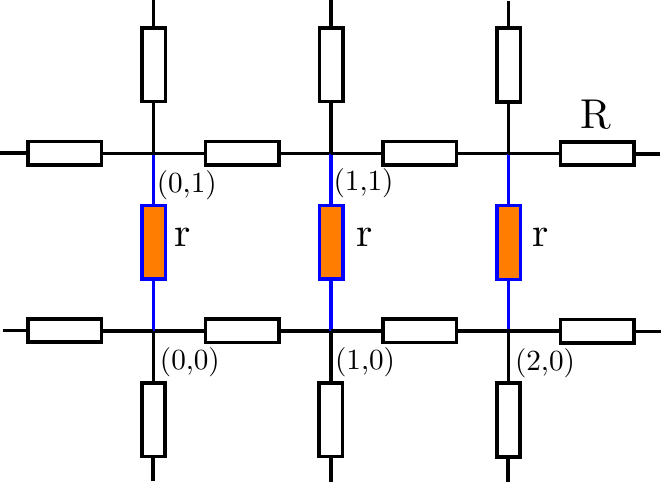}

\caption {\label{fig:horizontal} \footnotesize{\textbf{Perpendicular line defect in an infinite resistor lattice (modification on vertical edges).} A subset of the resistors, originally of resistance $R$ on the vertical edges, have been replaced by resistors $r$ along a horizontal line, as indicated. Several node coordinates $(x,y)$ are also labeled.}}
\end{figure}

In this chapter, we examine the case of a perpendicular line defect, when the network is modified as shown in figure~\ref{fig:horizontal}. Specifically, for each value of $x$, the resistance between the nodes $(x,0)$ and $(x,1)$ are changed from $R$ to $r$. Analogously to the previous chapter, this perturbation can be described by a linear operator $\hat{L}_1$, which takes the following form:
\begin{equation}
\hat{L}_1 = \sum_{x = 1}^{N_x} g   (|x,0\rangle-|x,1\rangle)(\langle x,0|-\langle x,1|) .
\end{equation} Since the system has translational symmetry in the horizontal direction, it is advantageous to switch to the reciprocal space in the $x$ direction. In the basis, defined by equation~\eqref{eq:basis}, the perturbation $\hat{L}_1$ becomes block-diagonal:
\begin{equation}
\hat{L}_1 = \sum_{k_x \in \mathcal{B}_x} |k_x\rangle\langle k_x| \otimes \hat{L}_1(k_x) ,
\label{eq:perturbation_perpendicular}
\end{equation} where each block is a dyadic product:
\begin{equation}
\hat{L}_1(k_x)=g(| 0\rangle-|1\rangle)(\langle 0|-\langle1|) .
\end{equation} Therefore, we can apply the Sherman–Morrison formula to calculate the Green's function of the perturbed system,
\begin{equation}
    \hat{G}(k_x) = \hat{G}_0(k_x) + \frac{g  \hat{G}_0(k_x) (| 0\rangle-|1\rangle)(\langle 0|-\langle1|) \hat{G}_0(k_x)}{1 - g  (\langle 0|-\langle 1|) \hat{G}_0(k_x) (| 0\rangle-|1\rangle)} .
    \label{eq:sm_perpendicular_1}
\end{equation}
The matrix elements of the Green's function, defined in equation~(\ref{eq:Green_basis}) are given by
\begin{equation}
    \begin{aligned}
    G(k_x,y_1,y_2) &= G_0(k_x,y_1,y_2) \\ &+ \frac{g \left[G_0(k_x,y_1,0)-G_0(k_x,y_1,1)\right] \left[G_0(k_x,y_2,0)-G_0(k_x,y_2,1)\right]}{1 - 2g  \left(G_0(k_x,0,0)-G_0(k_x,0,1)\right)} .
    \label{eq:sm_perpedicularl_2}    
    \end{aligned}
\end{equation} Using again equation~\eqref{eq:thermo_lim} for the limit $N_{x(y)} \to \infty$ and equation~\eqref{eq_perfect_green} for the matrix elements of $\hat{G}_0$ in the case of the perfect lattice, we find that the matrix elements of the Green’s function $\hat{G}$ are
\begin{equation}
\begin{aligned}
    G(k_x,y_1,y_2) &= G_0(k_x,y_1,y_2) \\ &+ g \frac{\euler^{-(|y_1| + |y_2|)s}}{(1+\euler^s)(1+\euler^s-2g)}\cdot \begin{cases}
\euler^{2s} &\text{if $y_1\geq1 \textrm{ and } y_2\geq1$},\\
1 &\text{if $y_1\leq0 \textrm{ and } y_2\leq0$},\\
(-\euler^s) &\text{other cases.}
\end{cases} 
\end{aligned}
\label{eq_green_perpendicular}
\end{equation}

Finally, we can transform the Green's function back to real space, and substitute this expression into equation~\eqref{eq_equivalent_resistance} to determine the effective resistance between points $(x_\mathrm{in},y_\mathrm{in})$ and $(x_\mathrm{out},y_\mathrm{out})$ within the perturbed lattice. The formula for the effective resistance, if $y_\textrm{in}\geq1 \textrm{ and } y_\textrm{out}\geq1$, becomes
\begin{equation}
    \begin{aligned}R(x&_\mathrm{out},y_\mathrm{out};x_\mathrm{in},y_\mathrm{in}) = R_0(x_\mathrm{out},y_\mathrm{out};x_\mathrm{in},y_\mathrm{in}) \\ &+ \frac{gR}{2\pi} \int_{-\pi}^\pi \mathrm{d}k_x\, \frac{\euler^{-2(y_\mathrm{out}-1)s} + \euler^{-2(y_\mathrm{in}-1)s} - 2 \cos\left[k_x(x_\mathrm{out}-x_\mathrm{in})\right]\euler^{-(y_\mathrm{out} + y_\mathrm{in}-2)s}}{(1+\euler^s)(1+\euler^s-2g)},
    \end{aligned}
\label{eq_resistance_perpendicular1}
\end{equation}
while for the case
$y_\textrm{in}\geq1 \textrm{ and } y_\textrm{out}\leq0$, it is
\begin{equation}
    \begin{aligned}
        R(x_\mathrm{out},y_\mathrm{out}&;x_\mathrm{in},y_\mathrm{in}) = R_0(x_\mathrm{out},y_\mathrm{out};x_\mathrm{in},y_\mathrm{in}) \\ &+ \frac{gR}{2\pi} \int_{-\pi}^\pi \mathrm{d}k_x\, \frac{\euler^{2y_\mathrm{out}s} + \euler^{-2(y_\mathrm{in}-1)s} + 2 \cos\left[k_x(x_\mathrm{out}-x_\mathrm{in})\right]\euler^{(y_\mathrm{out} - y_\mathrm{in}+1)s}}{(1+\euler^s)(1+\euler^s-2g)}.
    \end{aligned}
    \label{eq_resistance_perpendicular2}
\end{equation}
All other cases are equivalent to equations~\eqref{eq_resistance_perpendicular1} and \eqref{eq_resistance_perpendicular2} due to the horizontal mirror symmetry of the perturbed system, which implies that
\begin{equation}
R(x_\mathrm{out},y_\mathrm{out};x_\mathrm{in},y_\mathrm{in})=R(x_\mathrm{out},1-y_\mathrm{out};x_\mathrm{in},1-y_\mathrm{in}).
\end{equation}

\subsection{Analytical results for special $g$ values}

In this section, we investigate the special values of the perturbation parameter $g$, where the resulting resistance in the case of a perpendicular line defect can be reduced to that of a perfect square lattice. Excluding the trivial case ($g = 0$), we consider three distinct limiting cases: (i) $g = 1$, where the resistors along the defect line are completely removed; (ii) $g \rightarrow -\infty$, where the resistors along the defect line are short-circuited; and (iii) $g=1/2$, where the line defect consists of resistors with resistance $2R$.

In the first case ($g = 1$), the square lattice completely splits into two independent semi-infinite lattices. If a current is injected on one side of the defect, it does not influence the potential on the other side. Consequently, the Green's function vanishes when the $y$-coordinates of the two points lie on opposite sides of the line defect. If we are interested in the potential on the same side of the defect as the point of current injection, we can apply the image method described in \cite{8821548, 9078361}.  Consider a perfect square lattice in which the same currents are injected at the nodes with coordinates $(x_1, y_1)$ and $(x_1, 1 - y_1)$. Due to the symmetry of the system, no current flows through the resistors connecting nodes along the lines $y = 0$ and $y = 1$. Therefore, these resistors can be removed without changing the current or voltage distribution in the lattice. By removing them, we get a perpendicular line defect corresponding to the case $g = 1$. The Green’s function for this configuration can thus be written, based on the previous considerations, as:
\begin{equation}
G(x_1,y_1;x_2,y_2)=\begin{cases}
0 &\text{if $y_1\geq1 \textrm{ and } y_2\leq0$},\\
G_0(x_1,y_1;x_2,y_2)+G_0(x_1,1-y_1;x_2,y_2)&\text{if $y_1\geq1\textrm{ and } y_2\geq1$}.
\end{cases}
\end{equation}
Note that these can also be obtained by substituting the value $g=1$ into equation~(\ref{eq_green_perpendicular}).
From the Green’s function, the effective resistance can now be expressed in terms of the resistances measured in the perfect square lattice,
\begin{equation}
    \begin{aligned}
        R(x_\mathrm{out},y_\mathrm{out};x_\mathrm{in},y_\mathrm{in}) &= R_0(x_\mathrm{out}-x_\mathrm{in},y_\mathrm{out}-y_\mathrm{in}) - R_0(x_\mathrm{out}-x_\mathrm{in},y_\mathrm{out}+y_\mathrm{in}-1)\\ &+ \frac12 R_0(0,2y_\mathrm{in}-1) + \frac12 R_0(0,2y_\mathrm{out}-1).
    \end{aligned}
\end{equation}
For a few special pairs of points, the effective resistance can be derived exactly using table~\ref{perfect:table},
\begin{equation}
    \label{eq:res_perp_g=1}
    \begin{aligned}
    R(1,1;0,1) &= R_0(1,1)=\frac{2}{\pi},\\
    R(2,1;0,1) &= R_0(0,2) + R_0(1,2) -R_0(1,0) = 1,\\
     R(0,2;0,1) &= \frac{1}{2}R_0(0,1)+R_0(0,2)-\frac{1}{2}R_0(0,3) = \frac{8}{\pi} - 2.\\
   \end{aligned}
\end{equation}

By taking the limit $g\rightarrow-\infty$ in equation~(\ref{eq_green_perpendicular}), we obtain a Green’s function with a structure similar to that found for a parallel line defect with perturbation strength $g=-1$. The duality between these two cases becomes evident upon closer investigation. Specifically, when the perpendicular line defect consists of short circuits, the resistor between the points $(x,y=1)$ and $(x+1,y=1)$, as well as another between $(x,y=0)$ and $(x+1,y=0)$, are connected in parallel. This parallel connection can be replaced by a single resistor of resistance $R/2$. Performing this substitution for all values of $x$ transforms the configuration into a case of a parallel line defect with perturbation strength $g=-1$. The corresponding effective resistance of this parallel line defect case is denoted by $R^{||}(x_\mathrm{out},y_\mathrm{out};x_\mathrm{in},y_\mathrm{in})$, and its expression is given in equation~(\ref{eq_resistor_paralel_min1}). In the case of the perpendicular line defect with $g\rightarrow-\infty$, the resistances can be expressed as follows:
\begin{equation} 
R^\perp(x_\mathrm{out},y_\mathrm{out};x_\mathrm{in},y_\mathrm{in})=\begin{cases}
R^{||}(x_\mathrm{out},y_\mathrm{out}-1;x_\mathrm{in},y_\mathrm{in}-1)  &\text{if $y_\textrm{in}\geq1 \textrm{ and } y_\textrm{out}\geq1$},\\
R^{||}(x_\mathrm{out},y_\mathrm{out};x_\mathrm{in},y_\mathrm{in}-1)  &\text{if $y_\textrm{in}\geq1 \textrm{ and } y_\textrm{out}\leq0$},\\
R^{||}(x_\mathrm{out},y_\mathrm{out}-1;x_\mathrm{in},y_\mathrm{in})  &\text{if $y_\textrm{in}\leq0 \textrm{ and } y_\textrm{out}\geq1$},\\
R^{||}(x_\mathrm{out},y_\mathrm{out};x_\mathrm{in},y_\mathrm{in})  &\text{if $y_\textrm{in}\leq0 \textrm{ and } y_\textrm{out}\leq0$}.
\end{cases}
\label{eq:res_perp_g=-inf}
\end{equation}

Finally, in the last case, if we substitute $g=1/2$ into equation~(\ref{eq_green_perpendicular}), we obtain a Green's function that resembles the one found for a parallel line defect with $g=1$. The explanation for this duality is as follows. When $g=1$ in the parallel line defect, meaning the resistors along the defect are completely removed, then for each $x$, the resistors between $(x,y=-1)$ and $(x,y=0)$, as well as between $(x,y=0)$ and $(x,y=1)$, are connected in series. Replacing each such series connection with a single resistor of resistance $2R$ creates exactly the configuration of the perpendicular line defect, which is under our current consideration. Applying this duality, the effective resistances in the perpendicular line defect with $g=1/2$ can be directly mapped to those of the parallel defect case with $g=1$,
\begin{equation} 
R^\perp(x_\mathrm{out},y_\mathrm{out};x_\mathrm{in},y_\mathrm{in})=\begin{cases}
R^{||}(x_\mathrm{out},y_\mathrm{out};x_\mathrm{in},y_\mathrm{in})  &\text{if $y_\textrm{in}\geq1 \textrm{ and } y_\textrm{out}\geq1$},\\
R^{||}(x_\mathrm{out},y_\mathrm{out}-1;x_\mathrm{in},y_\mathrm{in})  &\text{if $y_\textrm{in}\geq1 \textrm{ and } y_\textrm{out}\leq0$},\\
R^{||}(x_\mathrm{out},y_\mathrm{out};x_\mathrm{in},y_\mathrm{in}-1)  &\text{if $y_\textrm{in}\leq0 \textrm{ and } y_\textrm{out}\geq1$},\\
R^{||}(x_\mathrm{out},y_\mathrm{out}-1;x_\mathrm{in},y_\mathrm{in}-1)  &\text{if $y_\textrm{in}\leq0 \textrm{ and } y_\textrm{out}\leq0$},
\end{cases}
\label{eq:res_perp_g=1/2}
\end{equation}
where now $R^{||}(x_\mathrm{out},y_\mathrm{out};x_\mathrm{in},y_\mathrm{in})$ is the effective resistance for a parallel line defect with $g=1$,which is given by  equation~(\ref{eq:resistance_g1}).

\subsection{Analytical results for special node coordinates}

In the presence of a perpendicular line defect, the total resistance was expressed by an integral formula, as given in equations~(\ref{eq_resistance_perpendicular1}) and (\ref{eq_resistance_perpendicular2}).  When the effective resistance is calculated between two nearby points that are also located close to the line defect, the integrals can be evaluated analytically. As an example, we calculate the effective resistance as a function of the perturbation strength between the points (0,0) and (0,1), as well as between (0,1) and (1,1),
\begin{equation}
    \begin{aligned}
        R(0,0;0,1) &=  \frac{1}{2-4g} \left[1-2F^\perp(g)\right],\\ 
           R(0,1;1,1) &=  \frac{1}{2}-\frac{g(4-8g+\pi g)}{2\pi(1-2g)^2} +\frac{(g-1)^2}{(1-2g)^2}F^\perp(g),
    \end{aligned}
    \label{resistance_analytical}
\end{equation}
where
\begin{equation} 
F^\perp(g)=\begin{cases}
\frac{2g\cdot\textrm{arccosh}\left(\frac{g}{\sqrt{2}(1-g)}\right)}{\pi\sqrt{(4-g)g - 2}} &\text{if $g\geq2-\sqrt{2}$},\\
\frac{2g\pi-2g\cdot\textrm{arccos}\left(\frac{g}{\sqrt{2}(g-1)}\right)}{\pi\sqrt{2 + (g - 4)g}} &\text{if $g< 2-\sqrt{2}$}.
\end{cases}
\end{equation}

\begin{table}[b]
	\caption{\label{R0:table2}
			Resistance of an infinite square lattice in the presence of a perpendicular line defect in units of $R$.}
	\begin{indented}
		\item[]\begin{tabular}{@{}c|c|c}
			\br
			$g$ & $R(0,0;0,1)$ & $R(0,1;1,1)$ \\
			\mr
            $-\infty$ & $0$ & $\frac{1}{\pi}$ \\[2ex]
            $-\frac{1}{\sqrt{2}-1}$ & $-\frac{1}{18}(-3+2\sqrt{2})(9+8\sqrt{3})$ & $\frac{-1+\sqrt{2}}{9\pi}(18+9\pi+8\sqrt{3}\pi-8\sqrt{6}\pi)$ \\[2ex]
			
            $0$ & $\frac12$ & $\frac12$ \\[2ex]
            $\frac{1}{\sqrt{2}+1}$ & $-\frac{1}{18}(3+2\sqrt{2})(-9+4\sqrt{3})$ & $\frac{1+\sqrt{2}}{9\pi}(-18-9\pi+4\sqrt{3}\pi+4\sqrt{6}\pi)$ \\[2ex]
            
            $1/2$ & $2-\frac4\pi$ & $\frac32-\frac3\pi$ \\[2ex]
            $1$ & $\infty$ & $ \frac{2}{\pi}$ \\[2ex]
			\br
		\end{tabular}
	\end{indented}
\end{table}

\begin{figure}[b]
\centering
\includegraphics[width=0.6\columnwidth]{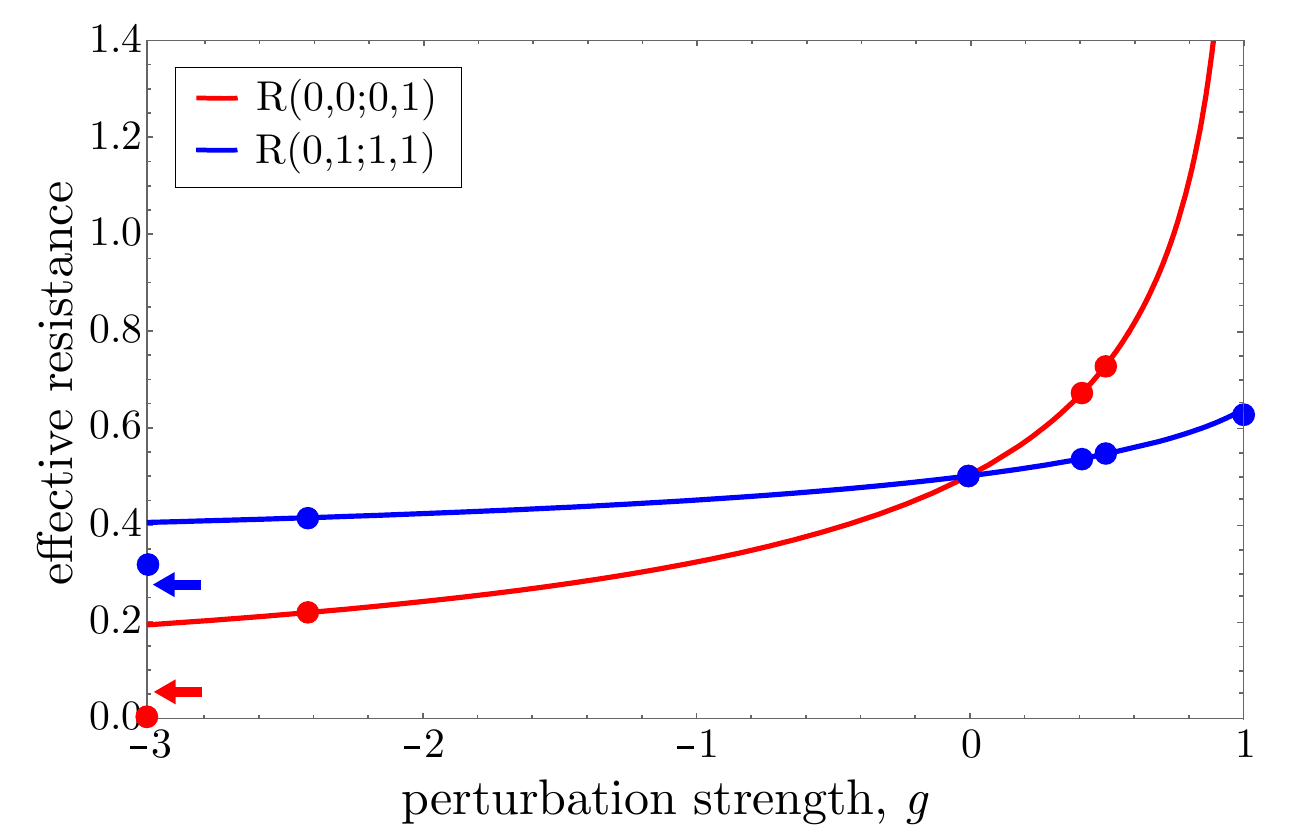}
\caption {\label{fig:perpendicular_result1} \footnotesize{\textbf{Effective resistances in the presence of a perpendicular line defect in an infinite resistor network} The red curve shows the dependence of the effective resistance on $g$ between the points $(0,0)$ and $(0,1)$, while the blue curve corresponds to the resistance between $(0,1)$ and $(1,1)$. The values listed in table \ref{R0:table2} are indicated by discrete markers. On the left edge of the plot, the limiting values, $g\rightarrow-\infty$, of the effective resistance are also marked.}}
\end{figure}

For several specific values of the perturbation strength $g$, we provide the exact values of the effective resistance in table \ref{R0:table2}. The values of the effective resistances are in agreement with the observations made in equations~\eqref{eq:res_perp_g=1}, \eqref{eq:res_perp_g=-inf}, \eqref{eq:res_perp_g=1/2}. The dependence of the effective resistance on the parameter $g$ is plotted in figure~\ref{fig:perpendicular_result1}.

\subsection{Numerical results for the effective resistances}

\begin{figure}[b]
\centering
\includegraphics[width=0.6\columnwidth]{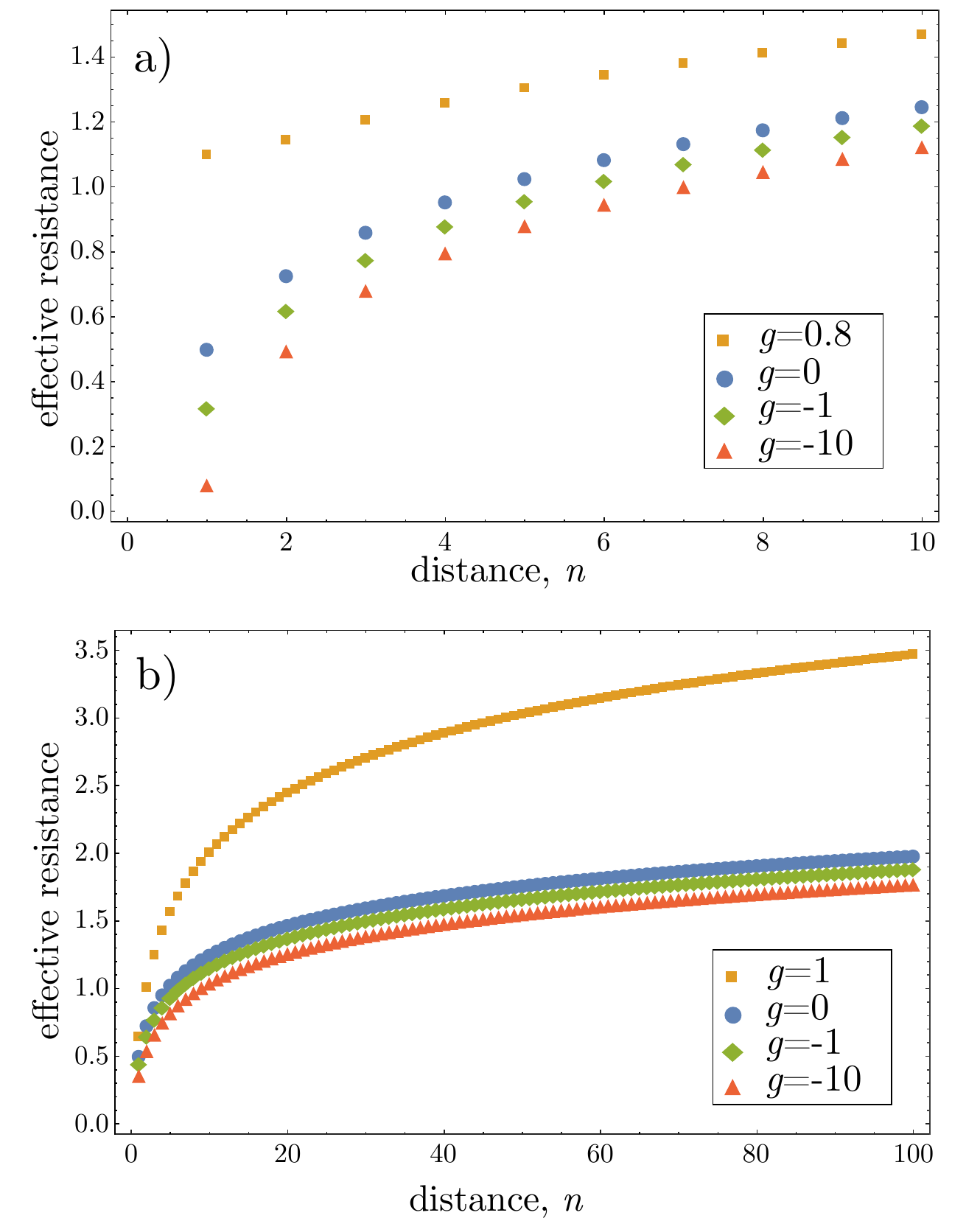}
\caption {\label{fig:perpendicular_figure6} \footnotesize{\textbf{Effective resistances in the presence of a perpendicular line defect in an infinite resistor network} a) The resistance $R(0,0;0,n)$, b) the resistance $R(0,1;n,1)$ in units of $R$ for different strength of the perturbation $g$.}}
\end{figure}

By numerically evaluating the integrals in equations~\eqref{eq_resistance_perpendicular1} and \eqref{eq_resistance_perpendicular2}, the effective resistance between any two points can be determined for a perpendicular line defect and for an arbitrary value of $g$. To demonstrate this, we consider two different configurations. In the first case, see figure~\ref{fig:perpendicular_figure6}a, one of the measurement points is fixed near the line defect at the position $(0,0)$, while the other is varied along the $y$ axis. The coordinates of the other measurement point are denoted by $(0,n)$. As expected, the effective resistance increases with the distance. Moreover, higher values of $g$ result in larger effective resistances, since the resistances within the line defect are also increased. It is also worth noting that for $g=1$, the effective resistance diverges. It occurs because the two measurement points lie in separate resistor networks that are not connected. In figure~\ref{fig:perpendicular_figure6}b, we plot the effective resistance between two points located along the line defect, with coordinates $(0,1)$ and $(n,1)$. It can be observed that as the distance between the points increases, the effective resistance increases in a logarithmic fashion.

\section{Alternating perpendicular line defect}

In the previous sections, we demonstrated that by applying the Sherman–Morrison formula from equation~\eqref{eq:Sherman-Morrison} on the Green’s functions expressed in momentum space, the effective resistance can be calculated for both parallel and perpendicular line defects. This general method remains applicable even when the system is perturbed in other ways along a line. 

\begin{figure}[h!]
\centering
\includegraphics[width=0.8\columnwidth]{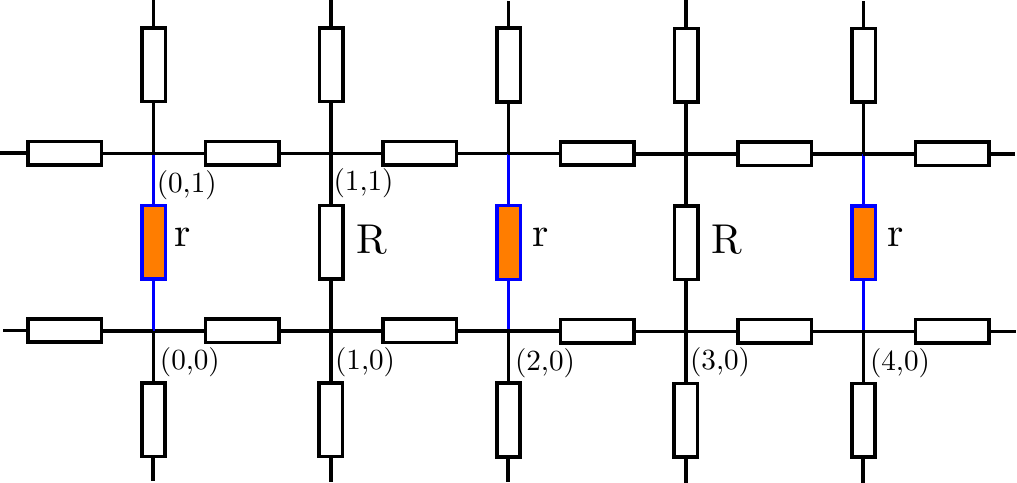}
\caption {\label{fig:alternating_perpendicular} \footnotesize{\textbf{Alternating perpendicular line defect in an infinite resistor lattice.} Setup, where only the resistance of every $n$th vertical link along a horizontal line is modified. Here, specifically $n=2$.}}
\end{figure}

In this chapter, we consider the alternating perpendicular line defect, which is a generalization of the perpendicular defect discussed in the previous section, where only the resistance of every $n$th link along the horizontal line is modified. A specific case with $n=2$ is shown in figure~\ref{fig:alternating_perpendicular}. The notations $R$, $r$, and $g = 1 - R/r$ used in this chapter are the same as those used previously.

Similarly as in equation~\eqref{eq:perturbation_perpendicular}, the perturbation operator $\hat{L}_1$ can be written as:
\begin{equation}
    \hat{L}_1 = g\sum_{x=1}^{N_x/n}\left(|nx,0\rangle - |nx,1\rangle\right) \left(\langle nx,0| - \langle nx,1|\right) .
    \label{eq:perturbation_alternating}
\end{equation} Using the partial basis transformation for the coordinate $x$ defined by equation~\eqref{eq:basis}, we obtain the following:
\begin{equation}
    \hat{L}_1 = \frac{g}{N_x} \sum_{x = 1}^{N_x/n} \sum_{k_x \in \mathcal{B}_x} \sum_{q_x \in \mathcal{B}_x} \euler^{\iu (q_x - k_x)nx}\left(|k_x,0\rangle - |k_x,1\rangle\right) \left(\langle q_x,0| - \langle q_x,1|\right) .
\end{equation} The summation over $x$ can be performed using the geometric series relation
\begin{equation}
    \sum_{x = 1}^{N_x/n} \euler^{\iu (q_x - k_x)nx} = \frac{N_x}{n}\sum_{j = 1}^n \delta_{q_x,k_x \oplus 2\pi j/n} \,,
    \label{eq:geometric_sum}
\end{equation} where $\oplus$ denotes addition modulo $2\pi$ as required by the properties of the Brillouin zone. After further simplifications, the final result yields:
\begin{equation}
    \hat{L}_1 = \frac{g}{n} \sum_{k_x \in \mathcal{B}_x} \sum_{j=1}^n \left(|k_x,0\rangle - |k_x,1\rangle\right) \left(\langle k_x \oplus 2\pi j/n,0| - \langle k_x \oplus 2\pi j/n,1|\right) .
\end{equation} One can notice that, unlike the perturbation operator expressed in equations~\eqref{eq_pertubation_blockform} and \eqref{eq:perturbation_perpendicular}, this one is not block-diagonal on the chosen basis. Therefore, it is convenient to write vectors $|k_x\oplus 2\pi j/n,0\rangle$ in a direct product form $|k_x\rangle \otimes |0\rangle_j$ for $-\pi/n \le k_x < \pi/n$ and each $j \in \{1,2,...,n\}$. This reduced set of $k_x$ vectors is denoted by $\mathcal{B}_x^n$. With this notation, the perturbation operator can be written in the following form:
\begin{equation}
    \hat{L}_1 = \frac{g}{n} \sum_{k_x \in \mathcal{B}_x^n} \sum_{l=1}^n \sum_{j=1}^n |k_x\rangle \langle k_x| \otimes \left(|0\rangle_l - |1\rangle_l\right) \left(\langle 0|_j - \langle 1|_j\right) .
\end{equation} In conclusion, by reducing the one-dimensional Brillouin zone from $\mathcal{B}_x$ to $\mathcal{B}_x^n$, we recovered the block-diagonal structure (with respect to different $k_x$ subspaces) of the perturbation operator. This aligns with our physical intuition: the alternating defect breaks down the translational invariance of the system such that in the $x$ direction, only translations by multiples of $n$ remain symmetries. 

In analogy to equations~\eqref{eq_pertubation_blockform} and \eqref{eq:perturbation_perpendicular}, this block-diagonal form allows us to introduce block operators as
\begin{equation}
    \hat{L}_1(k_x) = \frac{g}{n} \sum_{l=1}^n \left(|0\rangle_l - |1\rangle_l\right) \sum_{j=1}^n \left(\langle 0|_j - \langle 1|_j\right) .
\end{equation}Similarly, for the Laplacian of the perfect lattice, we have
\begin{equation}
    \hat{L}_0(k_x) = \sum_{k_y \in \mathcal{B}_y} \sum_{j=1}^n \left[2\cos \left(k_x + \frac{2\pi j}{n}\right) + 2\cos k_y - 4\right] \, |k_y\rangle_j \langle k_y|_j .
\end{equation} Note that these operators are not exactly the same as those used in Section \ref{chapter:parallel}, since here the values $k_x$ come from the reduced Brillouin zone $\mathcal{B}_x^n$ corresponding to the partially broken translational symmetries of the alternating defect. 
 
Using the Sherman--Morrison formula once again, we can obtain the following relation for the block operators of Green's function in this reduced Brillouin zone:
\begin{equation}
    \hat{G}(k_x) = \hat{G}_0(k_x) + \frac{g \sum_{l=1}^n \sum_{j=1}^n \hat{G}_0(k_x) \left(|0\rangle_l - |1\rangle_l\right) \left(\langle 0|_j - \langle 1|_j\right) \hat{G}_0(k_x) }{n - g \sum_{j=1}^n \sum_{l=1}^n \left(\langle 0|_j - \langle 1|_j\right)\hat{G}_0(k_x) \left(|0\rangle_l - |1\rangle_l\right) } .
\end{equation} Transforming this back to real space using a relation equivalent to equation \eqref{eq:basis}:
\begin{equation}
    |k_x \oplus 2\pi j/n,y\rangle = |k_x\rangle \otimes |y\rangle_j =\frac{1}{\sqrt{N_x}} \sum_{x = 1}^{N_x} \euler^{\iu k_xx} \euler^{2\pi\iu x j/n} |x,y\rangle , 
\end{equation} and applying the $N_{x(y)}\to\infty$ limit formula of equation~\eqref{eq:thermo_lim} leads to the following final result for the Green's function:
\begin{equation}
    \begin{aligned}
    G(x_1,y_1;&x_2,y_2) = G_0(x_1,y_1;x_2,y_2) \\ &+ \frac{g}{2\pi} \int_{-\pi/n}^{\pi/n}\mathrm{d}k_x\, \euler^{\iu k_x(x_1 - x_2)} \frac{ \sum_{j=1}^n \sum_{l=1}^n \euler^{2\pi \iu(x_1 j - x_2 l)/n} \Gamma_j(k_x,y_1) \Gamma_l(k_x,y_2)}{n - 2g \sum_{m=1}^n \Gamma_m(k_x,0)} ,
    \label{eq:alternating_Green}
    \end{aligned}
\end{equation} where we introduced the following notation for the difference of Green's functions:
\begin{equation}
    \Gamma_j(k_x,y) = G_0\left(k_x \oplus \frac{2\pi j}{n},y,0\right) - G_0\left(k_x \oplus \frac{2\pi j}{n},y,1\right) .
    \label{eq:Gamma}
\end{equation} The resistance can be derived from the Green’s function using equation~(\ref{eq_equivalent_resistance}), following the same procedure as in the previous cases.

\subsection{Numerical results for the effective resistances}

In the case of an alternating perpendicular line defect, as before, the Green’s function and consequently the effective resistance can be expressed as an integral formula. Even in the simplest configuration, the integral cannot be evaluated analytically; however, it is well-suited for numerical calculation, allowing us to determine the resulting resistances efficiently. To illustrate this, we provide a few examples below.

The table~\ref{R0:table3} contains the effective resistance between the points $(0,0)$ and $(0,1)$ if every $n$th link is removed from the network, i.e., $g = 1$. The measurement points are always the two endpoints of a removed edge. It can be observed that for $n = 1$, the lattice splits into two independent subsystems that are completely disconnected. In the limit $n \to \infty$, i.e., when only a single resistor is removed from the network, the effective resistance approaches the value $R$, as it is well known from references \cite{Gnädig_Honyek_Riley_2001,10.1119/1.1419104}.

\begin{table}[h]
	\caption{\label{R0:table3}
			Resistance between the $(0,0)$ and $(0,1)$ nodes in units of $R$ in the presence of an alternating perpendicular line defect, if every $n$th link is removed.}
	\begin{indented}
		\item[]\begin{tabular}{@{}c|c|c|c|c|c|}
			\br
			$n$ & 1 & 2 & 3 &5&10\\
			\mr
            
          R(0,0;0,1) & $\rightarrow\infty$ & 1.042 &  1.007& 1.001 & 1.000\\[2ex]
			\br
		\end{tabular}
	\end{indented}
\end{table}

For the case $n = 2$, the effective resistance was also determined using an independent numerical approach. Specifically, we applied the perturbative technique described in \cite{cserti2025generaltheoryperturbationinfinite}. 
We first consider a perfect square lattice from which the resistor between $(0,0)$ and $(0,1)$ is removed, resulting in an effective resistance $R$. If two additional resistors are removed symmetrically between $(-2,0)$ and $(-2,1)$ as well as between $(2,0)$ and $(2,1)$, the effective resistance increases to $1.035\,R$. Removing two more resistors between $(-4,0)$ and $(-4,1)$ and between $(4,0)$ and $(4,1)$ (five removed resistors in total) gives an effective resistance of $1.040\,R$. Continuing this procedure, after removing $13$ or more resistors we obtain $1.042\,R$. It is evident that the limit of this procedure corresponds exactly to an alternating perpendicular line defect, and therefore the effective resistance approaching $1.042\,R$ confirms the results of table~\ref{R0:table3}.

\section{Tilted line defect}
\label{sec:tilted}

In the previous section, we demonstrated that our general method for obtaining the lattice Green's function and equivalent resistances in the presence of line defects remains applicable even when the period of the defect is a multiple of the lattice constant. 

\begin{figure}[h!]
\centering
\includegraphics[width=0.8\columnwidth]{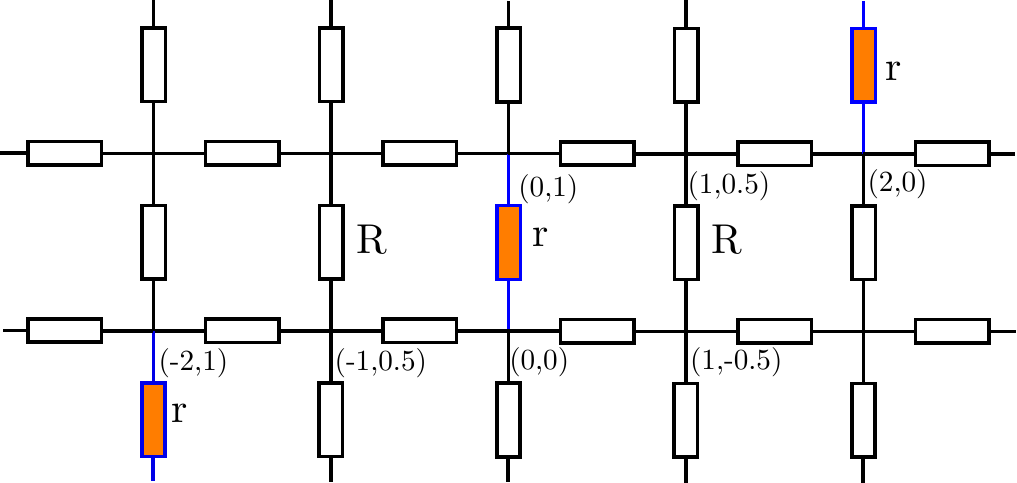}
\caption {\label{fig:tilted} \footnotesize{\textbf{Tilted line defect in an infinite resistor lattice.} Setup, where the links of modified resistance reside on a tilted line of rational slope $m/n$. Here, specifically $n=2$ and $m=1$.}}
\end{figure}

In this chapter, we consider an even more general case of the tilted line defect, where the resistances are modified periodically along an arbitrary, but non-vertical line. To characterize such a defect, we need two integer numbers: the horizontal ($n > 0$) and vertical ($m \ge 0$) components of the elementary translation vector pointing from a modified resistance to its nearest neighbor. A specific case with $n=2$ and $m=1$ is shown in figure~\ref{fig:tilted}. The notations $R$, $r$, and $g = 1 - R/r$ used in this chapter are the same as those used previously.

It turns out that the solution of this problem can be readily retraced to the case of the alternating perpendicular defect discussed in the previous section. Namely, we can relabel the real-space basis vectors as $|x,y\rangle \to |x,y - m x/n \rangle$. Note that the labels $y - mx/n$ are not necessarily integers. In this representation, the perturbation operator is formally the same as that in equation~\eqref{eq:perturbation_alternating}. As a consequence, all the derivations detailed in the previous section can be repeated. There is only one difference, the Laplacian of the perfect lattice is modified as
\begin{equation}
    \begin{aligned}
    \hat{L}_0 = \sum_{x = 1}^{N_x} \sum_{y = 1}^{N_y} \Big( &|x + 1,y - m/n\rangle \langle x,y| + |x - 1,y + m/n\rangle \langle x,y| \\ &+ |x,y + 1\rangle \langle x,y| + |x,y - 1\rangle \langle x,y| - 4\,|x,y\rangle \langle x,y| \Big) .
    \end{aligned}
\end{equation} Note that this is the same operator as in equation \eqref{eq:laplacian}, only the states have been relabeled. Using the definition of plane wave states, equation~\eqref{eq:plane_waves}, and summing up over the $x$ and $y$ variables, we can transform this into reciprocal space. The result reads:
\begin{equation}
    \hat{L}_0 = \sum_{k_x \in \mathcal{B}_x} \sum_{k_y \in \mathcal{B}_y} \left[2\cos\left(k_x - m k_y/n \right) + 2\cos k_y - 4\right] |k_x,k_y\rangle \langle k_x,k_y| .
\end{equation} The Green's operator $\hat{G}_0 = -\hat{L}_0^{-1}$ takes a similar form in reciprocal space:
\begin{equation}
    \begin{aligned}
        &\hat{G}_0 = \sum_{k_x \in \mathcal{B}_x} \sum_{k_y \in \mathcal{B}_y} G_0(k_x,k_y) |k_x,k_y\rangle \langle k_x,k_y| , \\
        &G_0(k_x,k_y) = \frac{1}{4 - 2\cos\left(k_x - m k_y/n \right) - 2\cos k_y} .
    \end{aligned}
\end{equation} Taking an inverse Fourier transform with respect to the second variable in the limit of equation~\eqref{eq:thermo_lim}, we obtain the real-reciprocal representation used before:
\begin{equation}
    G_0(k_x,y_1,y_2) = \frac{1}{4\pi} \int_{-\pi}^\pi\mathrm{d}k_y\,\frac{\euler^{\iu k_y(y_2-y_1)}}{2 - \cos\left(k_x - m k_y/n \right) - \cos k_y} .
\end{equation} Apart from a couple of special cases, this integral probably cannot be evaluated analytically. However, we can still calculate it numerically, and then substitution into equations~\eqref{eq:alternating_Green} and \eqref{eq:Gamma} yields the final result for the perturbed Green's function. 

The discussion above provides an efficient recipe for calculating the Green's function for a square lattice perturbed by a general tilted defect. This also allows us to compute the effective resistance between two arbitrary points of the lattice. However, it must be noted that the aforementioned relabeling of basis vectors breaks down when $n = 0$. This specific case must be handled differently, as presented in \ref{app:A}.


\section{Conclusions}

In this work, we computed the equivalent resistance between arbitrary nodes in a square lattice of resistors in the presence of various types of line defects. We began with simpler cases, such as perpendicular and parallel line defects, and then progressed to the more general tilted line defect, which includes the previous cases as special limits. A common feature of these systems is that the perturbation preserves translational invariance in one direction, allowing us to transform to reciprocal space along that axis and express the lattice Green’s function in this reciprocal basis. By taking into account the perturbation exactly using the Sherman–Morrison identity, we derived an analytical expression for the equivalent resistance. The resulting formula involves a one‑dimensional integral, which can be evaluated analytically in simpler cases and numerically for more complex configurations.

The proposed framework relies on the fact that the perturbation breaks translational symmetry only in a single direction. Therefore, it is generally applicable to describe line defects in arbitrary periodic 2D lattices (such as triangular or honeycomb lattices) and even 3D lattices (such as the cubic lattice). Furthermore, by including capacitive and inductive elements, the method can be generalized to networks with arbitrary complex impedances. This highlights its potential relevance to a wide range of applications, from classical resistor networks to topolectrical circuits. However, if the perturbation breaks translational symmetry in more than one independent direction,  for example, in the case of a semi-infinite line defect, the present approach is no longer applicable.

Although the description of line defects in resistor networks may seem like a purely theoretical problem, the underlying Poisson-type equation appears in many areas of physics. Our findings can also be applied to other classical systems, such as determining the stress field of dislocations in condensed matter. In addition, they are relevant for quantum systems, most notably topological insulators, where robust edge states emerge at the interface between regions with different topological invariants. The boundaries of two-dimensional topological systems can be effectively modeled by the kind of line defects we considered, implying that our framework could serve as a useful tool for describing edge states in lattice models governed by tight-binding Hamiltonians. This is the planned subject of one of our future publications.

We anticipate that our framework will prove to be a valuable tool for exploring both classical and quantum phenomena where periodic structures with linear-defect-type perturbations play a key role.

\section*{Acknowledgments}

We acknowledge fruitful discussions with Gyula Dávid. This research was supported
by the Ministry of Culture and Innovation, and the National Research, Development and Innovation Office within the Quantum Information National Laboratory of Hungary (Grant No. 2022-2.1.1-NL-2022-00004) and within the DKÖP-23 Doctoral Excellence Program, by the J\'{a}nos Bolyai Research Scholarship of the Hungarian Academy of Science, and by the NKFIH through the OTKA Grants FK 134437.

\appendix
\section{Alternating parallel line defect}
\label{app:A}

\begin{figure}[h!]
\centering
\includegraphics[width=0.8\columnwidth]{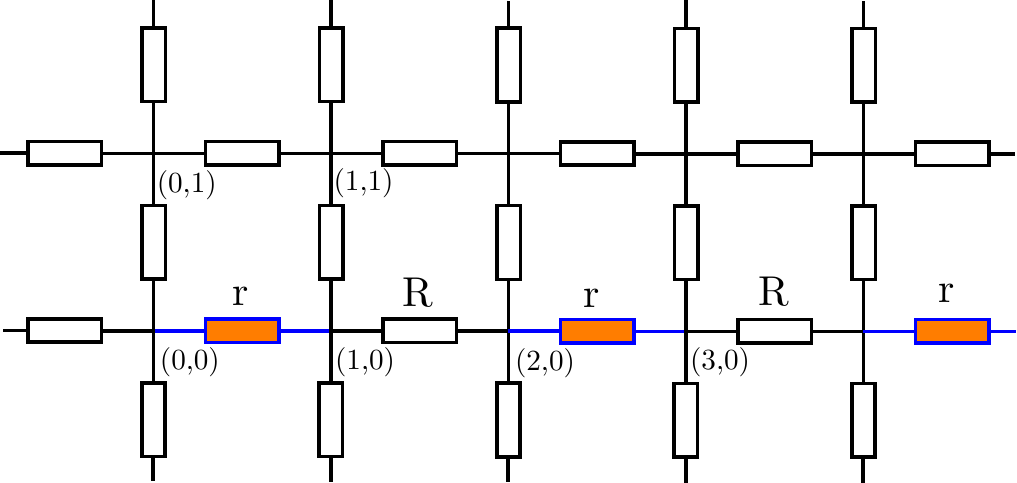}
\caption {\label{fig:alternating_parallel} \footnotesize{\textbf{Alternating parallel line defect in an infinite resistor lattice.} Setup, where only the resistance of every $n$th link along a horizontal line is modified. Here, specifically $n=2$.}}
\end{figure}

The only specific type of periodic line defect that is not covered by the general scheme for tilted defects presented in section~\ref{sec:tilted} is the alternating parallel line defect. This is a generalization of the parallel defect, where only the resistance of every $n$th link along a horizontal line is modified. A specific case with $n=2$ is shown in figure~\ref{fig:alternating_parallel}.

Let us start with the generalization of the perturbation operator for the alternating defect along the lines of equation~\eqref{eq_perturbation_parallel}:
\begin{equation}
    \hat{L}_1 = g\sum_{x=1}^{N_x/n}\left(2|nx,0\rangle\langle nx,0| - |nx,0\rangle\langle nx+1,0|-|nx+1,0\rangle\langle nx,0|\right) .
\end{equation} Using the partial basis transformation to reciprocal space defined by equation~\eqref{eq:basis}, we obtain the following:
\begin{equation}
    \hat{L}_1 = \frac{g}{N_x} \sum_{x = 1}^{N_x/n} \sum_{k_x \in \mathcal{B}_x} \sum_{q_x \in \mathcal{B}_x} \euler^{\iu (q_x - k_x)nx}\left( 2 - \euler^{\iu q_x} - \euler^{-\iu k_x} \right) |k_x,0\rangle \langle q_x,0| .
\end{equation} The summation over $x$ can be performed using the geometric series relation given in equation~\eqref{eq:geometric_sum}. After further simplifications, and separating vectors $|k_x\oplus 2\pi j/n,0\rangle$ in direct product form $|k_x\rangle \otimes |0\rangle_j$ like before, the perturbation operator becomes a block-diagonal matrix once again.  The corresponding block operators take the following form:
\begin{equation}
    \hat{L}_1(k_x) = \frac{2g}{n} \sum_{j=1}^n \sum_{l=1}^n \left[1 - \euler^{\iu \pi (j - l)/n} \cos \left(k_x + \frac{j + l}{n} \pi \right)\right] |0\rangle_l\langle 0|_j .
\end{equation} A new feature of this case is that the operators $\hat{L}_1(k_x)$ cannot be expressed as a dyadic product of vectors, but only a linear combination of dyads. Therefore, instead of the Sherman--Morrison formula, we must make use of the more general Sherman--Morrison--Woodbury identity \cite{Numerical_Recipes_3rd_10.5555:book, guttman1946enlargement, Woodbury:cikk} this time.

To this end, let us define a finite $n\times n$ matrix $\mathbf{C}(k_x)$ as a representation of the operator $\hat{L}_1(k_x)$ with respect to the basis $\{|0\rangle_1, |0\rangle_2, ..., |0\rangle_n\}$. The explicit definition is given through the following coefficients:
\begin{equation}
    C_{lj}(k_x) = \frac{2g}{n} \left[1 - \euler^{\iu \pi (j - l)/n} \cos \left(k_x + \frac{j + l}{n} \pi \right)\right] .
\end{equation} Furthermore, we can  define a vector $|\alpha\rangle$ as follows:
\begin{equation}
    |\alpha\rangle = \left( |0\rangle_1, |0\rangle_2, ..., |0\rangle_n \right) .
\end{equation} With these objects, the Sherman--Morrison--Woodbury identity can be readily applied, following reference~\cite{cserti2025generaltheoryperturbationinfinite}. The result for the block operators of the Green's function reads
\begin{equation}
    \hat{G}(k_x) = \hat{G}_0(k_x) + \hat{G}_0(k_x)|\alpha\rangle \left[\mathbf{C}^{-1}(k_x) - \langle\alpha |\hat{G}_0(k_x)|\alpha\rangle \right]^{-1} \langle\alpha | \hat{G}_0(k_x) .
\end{equation} The inversion of the $n\times n$ matrices appearing above can be performed numerically. From this closed formula, the real-space coefficients of the Green's function and the equivalent resistance can be expressed following the lines of the previous sections. 

\section*{References}
\bibliographystyle{iopart-num}
\bibliography{paper}

\end{document}